\documentclass[10pt,a4paper]{article}

\usepackage{epsfig}
\usepackage{amsmath}
\usepackage{txfonts}
\usepackage[round]{natbib}

\title{Networks and the Epidemiology of Infectious Disease}
\author{Leon Danon$^1$ \and Ashley P. Ford$^2$ \and Thomas House$^3$ \and Chris P. Jewell$^2$ \and
Matt J. Keeling$^{1,3}$ \and Gareth O. Roberts$^2$ \and Joshua V. Ross$^4$ \and Matthew C. Vernon$^1$}

\date{\begin{raggedright}
All authors contributed equally to this manuscript.\\
{\small
$^1$ Dept of Biological Sciences, University of Warwick, Coventry, CV4 7AL, UK.\\
$^2$ Dept of Statistics, University of Warwick, Coventry, CV4 7AL, UK.\\
$^3$ Warwick Mathematics Institute, University of Warwick, Coventry, CV4
      7AL, UK.\\ $^4$ School of Mathematical Sciences, University of Adelaide, SA 5005,
    Australia.\\ }
\end{raggedright}
}

\begin{document}

\maketitle

\section{Introduction}
The science of networks has revolutionised research into the dynamics of
interacting elements. The associated techniques have had a huge impact in a
range of fields, from computer science to neurology, from social science to
statistical physics. However, it could be argued that epidemiology has embraced
the potential of network theory more than any other discipline. There is an
extremely close relationship between epidemiology and network theory that dates
back to the mid-1980s \citep{Klovdahl:1985p1977, May:1987}. This is because the
connections between individuals (or groups of individuals) that allow an
infectious disease to propagate naturally define a network, while the network
that is generated provides insights into the epidemiological dynamics. In
particular, an understanding of the structure of the transmission network
allows us to improve predictions of the likely distribution of infection and
the early growth of infection (following invasion), as well as allowing the
simulation of the full dynamics. However the interplay between networks and
epidemiology goes further; because the network defines potential transmission
routes, knowledge of its structure can be used as part of disease control. For
example, contact tracing aims to identify likely transmission network
connections from known infected cases and hence treat or contain their contacts
thereby reducing the spread of infection.  Contact tracing is a highly
effective public health measure as it uses the underlying transmission dynamics
to target control efforts and does not rely on a detailed understanding of the
etiology of the infection. It is clear therefore that the study of networks and
how they relate to the propagation of infectious diseases is a vital tool to
understanding disease spread and therefore informing disease control.\\

Here we review the growing body of research concerning the spread of infectious
diseases on networks, focusing on the interplay between network theory and
epidemiology. The review is split into four main sections, which examine: the
types of network relevant to epidemiology; the multitude of ways these networks
can be characterised; the statistical methods that can be applied to either
infer the likely network structure or the epidemiological parameters on a
realised network; and finally simulation and analytical methods to determine
epidemic dynamics on a given network. Given the breadth of areas covered and
the ever-expanding number of publications (over seven thousand papers have been
published concerning infectious diseases and networks) a comprehensive review
of all work is impossible. Instead, we provide a personalised overview into the
areas of network epidemiology that have seen the greatest progress in recent
years or have the greatest potential to provide novel insights. As such
considerable importance is placed on analytical approaches and statistical
methods which are both rapidly expanding fields. We note that a range of other
network-based processes (such as the spread of ideas or panic) can be modelled
in a similar manner to the spread of infection, however in these contexts the
transmission process is far less clear; therefore throughout this review we
restrict our attention to epidemiological issues.\\

\section{Networks, Data and Simulations}

There are a wide number of network structures and types that have been
utilised when considering the spread of infectious diseases. Here, we
consider the most common forms and explain their uses and
limitations. Later, we review the implications of these structures for
the spread and control of infectious diseases.

\subsection{The Ideal Network}

We start our examination of network forms by considering the ideal
network that would allow us to completely describe the spread of any
infectious pathogen. Such a network would be derived from an
omniscient knowledge of individual behaviour. We define $G_{i,j}(t)$
to be a time-varying, real, high-dimensional variable that informs
about the strength of all potential transmission routes from
individual $i$ to individual $j$ at time $t$. Any particular
infectious disease can then be represented as a function
($f_{\mathrm{pathogen}}$) translating this high-dimensional variable
into an instantaneous probabilistic transmission rate (a single real
variable). In this ideal, $G$ subsumes all possible transmission
networks, from sexual relations to close physical contact,
face-to-face conversations, or brief encounters, and quantifies the
time-varying strength of this contact. The disease function then picks
out (and combines) those elements of $G$ that are relevant for
transmission of this pathogen, delivering a new (single-valued)
time-varying infection-specific matrix ($T_{i,j}(t) =
f_{\mathrm{pathogen}}(G_{i,j}(t))$).  This infection-specific matrix
then allows us to define the stochastic dynamics of the infection
process for a given pathogen. (For even greater generality, we may
want to let the pathogen-specific function $f$ also depend on the time
since an individual was infected, such that time-varying infectivity
or even time-varying transmission routes can be accommodated.)\\

Obviously, the reality of transmission networks is far from this
ideal.  Information on the potential transmission routes within a
population tends to be limited in a number of aspects. Firstly, it is
rare to have information on the entire population; most networks rely
on obtaining personal information on participants and therefore
participation is often limited. Secondly, information is generally
only recorded on a single transmission route (e.g.  face-to-face
conversation or sexual partnership) and often this is merely recorded
as the presence or absence of a contact rather than attempting to
quantify the strength or frequency of the interaction. Finally, data
on contact networks are rarely dynamic; what is generally recorded is
whether a contact was present during a particular period with little
consideration given to how this pattern may change over time. In the
light of these departures from the ideal, it is important to consider
the specifics of different networks that have been recorded or
generated, and understand their structure, uses and limitations.\\

\subsection{Realised Encounter Networks}

One of the few examples of where many of the potential transmission
routes within a population have been documented comes from the spread
of sexually transmitted infections (STIs). In contrast with airborne
infections, STIs have very obvious transmission routes --- sex acts
(or sharing needles during intravenous drug use) --- and as such these
potential transmission routes should be easily remembered (Figure
1A). Generally the methodology replicates that adopted during contact
tracing, getting an individual to name all their sexual partners over
a given period, these partners are then traced and asked for their
partners, and the process is repeated --- this is known as {\it
  snowball sampling} \citep{Goodman:1961} (Figure 1B). A related
methodology is {\it respondent driven sampling}, where individuals are
paid both for their participation and the participation of their
contacts while protecting each individual's anonymity  
\citep{Heckathorn:1997}. This approach, while  suitable for hidden 
and hard to reach populations, has a number of limitations, both practical and theoretical: recruiting 
people into the study, getting them to disclose such highly personal information,
imperfect recall from participants, the inability to find all partners, and the clustering of contacts. In
addition, there is the theoretical issue, that this algorithm will
only find a single connected component within the population, and it
is quite likely that multiple disjoint networks exist
\citep{Jolly:2001p8230}. \\

Despite these problems, and motivated by the desire to better
understand the spread of HIV and other STIs, several pioneering
studies were performed.  Probably the earliest is discussed by
\citet{Klovdahl:1985p1977} and utilises data collected by the Center
for Disease Control from 19 patients in California suffering from
AIDS, leading to a network of 40 individuals. Other larger-scale
studies have been performed in Winnipeg, Manitoba, Canada
\citep{Wylie:2001p3817} and Colorado Springs, Colorado, U.S.A.
\citep{Klovdahl:1994p3785}. In both of these studies, participants
were tested for STIs, and the distribution of infection compared to
the underlying network structure. Work done on both of these networks
has generally focused on network properties and the degree to which
these can explain the observed cases; no attempt was made to use these
networks predictively in simulations.  In addition, in the Colorado
Springs study tracing was generally only performed for a single
iteration, although many initial participants in high-risk groups were
enrolled; while in the Manitoba study tracing was performed as part of
the routine information gathered by public health nurses.  Therefore,
while both provide a vast amount of information on sexual contacts, it
is not clear if the results are truly a comprehensive picture of the
network and sampling biases may corrupt the resulting network
\citep{Ghani:1998p8627}. In addition, compared to the ideal network,
these sexual contact networks lack any form of temporal information,
instead they provide an integration of the network over a fixed time
period, and generally lack information on the potential strength of a
contact between individuals. Despite these difficulties, they continue
to provide an invaluable source of information on human sexual
networks and the potential transmission routes of STIs. In particular
they point to the extreme levels of heterogeneity in the number of
sexual contacts over a given period --- and the variance in the number
of contacts has been shown to play a significant role in early
transmission dynamics \citep{AndersonMay}\\

One of the few early examples of the simulation of disease
transmission on an observed network comes from a study of a small
network of 22 injection drug users and their sexual partners
\citep{Bell:1999} (Figure 1A). In this work the risk of transmission between two
individuals in the network was imputed based on the frequency and
types of risk behaviour connecting those two individuals. HIV
transmission was modelled using a monthly time-step and single index
case, and simulations were run for varying lengths of (simulated)
time. This enabled a node's position in the network (as characterised
by a variety of measures) to be compared with how frequently it was
infected during simulations, and how many other nodes it was typically
responsible for infecting.\\

A different approach to gathering social network and behavioural data
was initiated by the Human Dynamics group at MIT and illustrates how
modern technology can assist in the process of determining
transmission networks. One of the first approaches was to take
advantage of the fact that most people carry mobile phones
\citep{eagle2006reality}. In 2004, 100 Nokia 6600 smart-phones
pre-installed with software were given to MIT students to use over the
course of the 2004--05 academic year. Amongst other things, data were
collected using Bluetooth to sense other mobile phones in the
vicinity. These data gave a highly detailed account of individuals
behaviour and contact patterns. However, a limitation of this work was
that Bluetooth has a range of up to 25 meters, and as such networks
inferred from these data may not be epidemiological meaningful.\\

A more recent study into the encounters between wild Tasmanian devils
in the Narawntapu National Park in northern Tasmania utilised a
similar technological approach \citep{Hamede:2009p7045}. In this work
46 Tasmanian devils were fitted with proximity loggers, that could
detect and record the presence of other loggers within a 30cm
range. As such these loggers were able to provide detailed temporal
information on the potential interaction between these 46 animals.
This study was initiated to understand the spread of Tasmanian devil
facial tumour disease, which causes usually-fatal tumours that can be
transmitted between devils if they fight and bite each other. Although
only 27 loggers with complete data were recovered, and although the
methodology only recorded interaction between the 46 Devils in the
study, the results were highly informative (generating a network that
was far from random, heterogeneous and of detailed temporal
resolution). Analyses based on the structure of this network suggested
that targeted measures, that focus on the most highly connected ages
or sex, were unlikely to curtail the spread of this infection. Of
perhaps greater relevance is the potential this method illustrates for
determining the contact networks of other species (including humans)
--- the only limitation being the deployment of a suitable number of
proximity loggers.\\

\subsection{Inferred Encounter Networks} \label{IEN_subsect}

Given the huge logistical difficulties of capturing the full network
of interactions between individuals within a population, a variety of
methods have been developed to generate synthetic networks from known
attributes. Generally such methods fall into two classes: those that
utilise egocentric information, and those that attempt to simulate the
behaviour of individuals.\\

Egocentric data generally consists of information on a number of
individuals (the egos) and their contacts (the alters). As such the
information gathered is very similar to that collected in the sexual
contact network studies in Manitoba and Colorado Springs, but with
only the initial step of the snowball sampling was performed; the
difference is that for the majority of egocentric data the identity of
partners (alters) is unknown and therefore connections between egos
cannot be inferred (Figure 1C). The data therefore exists as multiple
independent `stars' linking the egos to the alters, which in itself
provides valuable information on heterogeneities within the
network. Two major studies have attempted to gather such egocentric
information: the NATSAL studies of sexual contacts in the UK
\citep{Johnson:1992p5390, Johnson:1994, Johnson:2001p5354,
  Copas:2002p8321}, and the POLYMOD study of social interactions
within 8 European countries \citep{Mossong:2008p5450}. The key to
generating a network from such data is to probabilistically assign
each alter a set of contacts drawn from the information available from
egos; in essence, using the ego data to perform the next step in the
snowball sampling algorithm. The simplest way to do this is to
generate multiple copies of all the egos and to consider the contacts
from each ego to be ``half-links"; the half-links within the network
can then be connected at random generating a configuration network
\citep{Molloy:1995, Molloy:1998, Read:2008p4987}; if more information
is available on the status (age, gender, etc.) of the egos and alters
then this can also be included and will reduce the set of half-links
that can be joined together. However, in the vast majority of
modelling studies, the egocentric data have simply been used to
construct WAIFW (who-acquires-infection-from-whom) matrices
\citep{Johnson:2001p5354, Mossong:2008p5450, Baguelin:2010p7690} that
inform about the relative levels of transmission between different
groups (e.g. based on sexual activity or age) but neglect the implicit
network properties. This matrix-based approach is often reliable: for
STIs it is the extreme heterogeneity in the number of contacts (which
are close to being power-law or scale-free distributed, see section
3.2) that drives the infection dynamics \citep{Liljeros:2001p5552}
although larger-scale structure does play a role \citep{Ghani:2000};
for social interactions it is the assortativity between (age-) groups
that controls the behaviour, with the number of contacts being
distributed as a negative binomial \citep{Mossong:2008p5450}.  The
POLYMOD matrices have therefore been extensively used in the study of
the H1N1 pandemic in 2009, providing important information about the
cost-effective vaccination of different age-classes
\citep{Medlock:2009p8088, Baguelin:2010p7690}.\\

\begin{figure}[p!]
\centerline{\includegraphics[width=1.1\columnwidth]{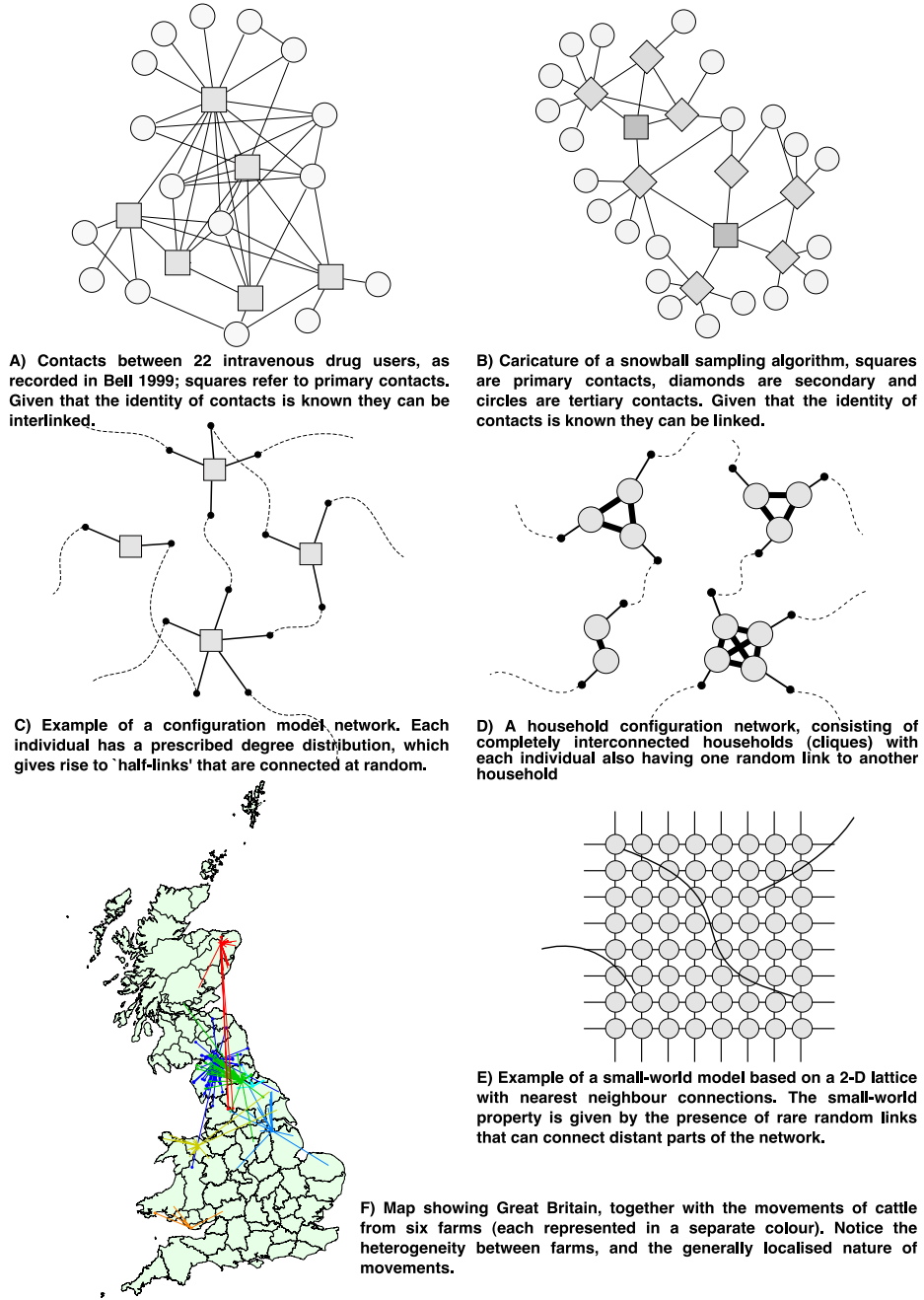}}
\caption{Examples of networks used in epidemiology}
\end{figure}

The general configuration model approach of randomly linking together
``half-links" from each ego \citep{Molloy:1995, Molloy:1998} has been
adopted and modified to consider the spread of STIs. In particular
simulations have been used to consider the important of concurrency in
sexual networks \citep{Kretzschmar:1996,Morris:1997}, where
concurrency is defined as being in two active sexual partnerships at
the same time. A dynamic sexual network was simulated, with
partnerships being broken and reformed such that the network density
remained constant over time. The likelihood of two nodes forming a
partnership depended on their degree, but this relationship could be
tuned to make concurrency more or less common, and to make the mixing
assortative or disassortative based on the degrees of the two nodes.
Transmission of an STI (such as gonorrhoea and chlamydia
\citep{Kretzschmar:1996} or HIV \citep{Morris:1997}) was then
simulated upon this dynamic network, showing that increasing
concurrency substantially increased the growth rate during the early
phase of an epidemic (and, therefore, its size after a given period of
time). This greater growth rate was related to the increase in giant
component size (see section 3.1) that was caused by increased
concurrency.\\

A slightly more general approach to the generation of model sexual
networks was employed by \citet{Ghani:1997}. In their network model,
individuals had a preferred number of concurrent partners and
duration of partnerships, and their level of assortativity was
tunable. A gonorrhoea-like infection was simulated on the resulting
dynamic network. Regression models were used to consider the
association between network structures (either snapshots of the state
of the network at the end of simulation, or accumulated over the last
90 days of simulation) and prevalence of infection. These simulations
showed that increasing levels of concurrent partnerships made invasion
of the network more likely, and also that the mixing patterns of the
most sexually active nodes were most important in determining the
final prevalence of infection within the
population~\citep{Ghani:1997}. The same model was later used to
consider the importance of different structural measures and sampling
strategies, showing that it was important to endeavour to identify
infected individuals with a high number of sexual partners in order to
correctly define the high-risk group for
interventions~\citep{Ghani:2000}.\\

The alternative approach of simulating the behaviour of individuals is
obviously highly complex and fraught with a great deal of
uncertainty. Despite these problems, three groups have attempted just
such an approach: Longini's group at Emory \citep{Halloran:2002p2396,
  Longini:2005p6498, Germann:2006p4899, Longini:2007}, Ferguson's
group at Imperial \citep{Ferguson:2005p6608, Ferguson:2006p78} and
Eubank's group at Los Alamos / Virginia Tech \citep{Chowell:2003p8430,
  Eubank:2004}. The models of both Longini and Ferguson are primarily
agent-based models, where individuals are assigned a home and work
location within which they have frequent infection-relevant contacts
together with more random transmission in their local
neighbourhood. The Longini models separate the entire population into
sub-units of 2000 individuals (for the USA) or 13000 individuals (for
South-East Asia) who constitute the local population where random
transmission can operate; in contrast the Ferguson models assign each
individual a spatial location and random transmission occurs via a
spatial kernel. In principle, both of these models could be used to
generate an explicit network model of possible contacts. The Eubank
model is also agent-based aiming to capture the movements of 1.5
million people in Portland, Oregon, USA; but these movements are then
used to define a network based on whether two individuals occur in the
same place (there are 180 thousand places represented in the model) at
the same time. It is this network that is then used to simulate the
spread of infection. While in principle this Eubank model could be
used to define a temporally varying and real-valued network (where the
strength of connection would be related to the type of mixing in a
location and the number of people in the location), in the
epidemiological publications \citep{Eubank:2004} the network is
considered as a static contact network in which extreme heterogeneity
in numbers of contacts is again predicted and the network has `small
world' like properties (see below). A similar approach of generating
artificial networks of individuals for stochastic simulations of
respiratory disease has been recently applied to influenza at the
scale of the United States, and the software made generally
available~\citep{Chao:2010}. This software took a more realistic
dynamic network approach and incorporated flight data within the
United States, but was sufficiently resource-intensive to require
specialist computing facilities (a single simulation taking around 192
hours of CPU time).  All three models have been used to consider
optimal control strategies, determining the best deployment of
resources in terms of limiting transmission associated with different
routes. The predicted success of various control strategies therefore
critically depends on the strength of contacts within the home, at
work, within social groups, and that occur at random.\\

Whilst smallpox has been eradicated, concern remains about the
possibility of a deliberate release of the disease. The stochastic
simulation models of the Longini group have predominantly focused on
methods of controlling this infection \citep{Halloran:2002p2396,
  Longini:2007}. Their early work utilised networks of two thousand people with
realistic age, household size and school attendance distributions,
with the likelihood of each individual becoming infected being derived
from the number and type of contacts with infectious individuals
\citep{Halloran:2002p2396}. This paper focused on the use of
vaccination to contain a small-scale outbreak of smallpox, and
concluded that early mass-vaccination of the entire population was
more effective than targeted vaccination if there was little or no
immunity in the population.  Later models \citep{Longini:2007}
combined these sub-networks of two thousand people into a larger network of
fifty thousand people (with one hospital), and the adult population were able
to contact each other through workplaces and high schools.  Here the
focus was on surveillance and containment which were generally
concluded to be sufficient to control an outbreak. The epidemiological
work of the Eubank group has also focused on a release of smallpox,
although these simulations showed that encouraging people to stay at
home as soon as they began to feel unwell was more important than
choice of vaccination protocol \citep{Eubank:2004}; this may in part
be attributed to the scale-free structure of the network and hence the
super-spreading nature of some individuals.  \\

The Ferguson models have primarily been used to consider the spread and
control of pandemic influenza, examining its potential spread from an
initial source in South-East Asia \citep{Ferguson:2005p6608}, and its
spread in mainland USA and Great Britain \citep{Ferguson:2006p78}. The
models of South-East Asia were primarily based on Thailand, and
included demographic information and satellite-based spatial measures
of population density. It focused on containment by the targeted use
of antiviral drugs and suggested that as long as the reproductive
ratio ($R_0$) of a novel strain was below 1.8 it could be contained by
the rapid use of targeted antivirals and social distancing. However,
such a strategy could require a stockpile of around 3 million
antiviral doses. The models based on the USA and Great Britain,
considered a wider range of control measures, including
school-closures, household prophylaxis using antiviral drugs, and
vaccination, and predicted the likely impact of different policies. \\

\subsection{Movement Networks} \label{Movements}

An alternative source of network information comes from the recorded
movements of individuals. Such data frequently describe a relatively
large network as information on movements is often collected by
national or international bodies. The network of movements therefore
has nodes representing locations (rather than individuals) and edges
weighted to capture the number of movements from one location to
another --- as such the network is rarely symmetric.  Four main forms
of movement network have played important roles in understanding the
spread of infectious diseases: the airline transportation network
\citep{Hufnagel:2004p4303, Guimera:2005p4438}, the movement of
individuals to and from work \citep{Hall:2007p3995, Viboud:2006p50}, the
movement of dollar bills (from which the movement of people can be
inferred) \citep{Brockmann:2006p8431}, and the movement of livestock
(especially cattle) \citep{Green:2006, Robinson:2007p4945}.  While the
structure of these networks has been analysed in some detail, to
develop an epidemiological model requires a fundamental assumption
about how the epidemic progresses within each locations. All the examples
considered in this section make the simplifying assumption that
the epidemic dynamics within each location are defined by random
(mean-field) interactions, with the network only informing about the flow of individuals or
just simply the flow of infection between populations --- such a
formulation is known as a metapopulation model
\citep{Hanski_book_2004}.\\

Probably the earliest work using detailed movement data to drive
simulations comes from the spread of 1918 pandemic influenza in the
Canadian Subarctic, based on records kept by the Hudson's Bay Company
\citep{Sattenspiel:1998}. A conventional SIR metapopulation model was
combined with a network model (the nodes being three fur trading posts
in the region: God's Lake, Norway House, and Oxford House) where some
individuals remained in their home locations whilst others moved
between locations, based on records of arrivals and departures
recorded in the post journals. Whilst this model described only a
small population, it was able to be parameterised in considerable
detail due to the quality of demographic and historical data
available, and showed that the movement patterns observed interacted
with the starting location of a simulated epidemic to change the
relative timings of the epidemics in the three communities, but not
the overall impact of the disease.\\

The movement of passenger aircraft as collated by the International
Air Transport Association (IATA) provides very useful information
about the long-distance movement of individuals and hence how rapidly
infection is likely to travel around the globe
\citep{Hufnagel:2004p4303, Colizza:2006p4294, Colizza:2007}. Unlike many other network models which are stochastic individual-level simulations, the
work of \citet{Hufnagel:2004p4303} and \citet{Colizza:2006p4294} was
based on stochastic Langevin equations (effectively differential
equations with noise included). The early
work by \citet{Hufnagel:2004p4303} focused on the spread of SARS, and
showed a remarkable degree of similarity between predictions and the
global spread of this disease. This work also showed that extreme
sensitivity to initial conditions arises from the structure of
the network, with outbreaks starting in different locations generating
very different spatial distributions of infection. The work of Colizza
was more focused towards the spread of H5N1 pandemic influenza arising
in South-East Asia, and its potential containment using antiviral
drugs.  However it was H1N1 influenza from
Mexico that initiated the 2009 pandemic, but again the IATA flight
data provided a useful prediction of the early spread
\citep{Khan:2009p8672, Balcan:2009}. While such global movement
networks are obviously highly important in understanding the early
spread of pathogens, they unfortunately neglect more localised
movements \citep{Viboud:2006p5270} and individual-level transmission
networks. However, recent work has aimed to overcome this first issue
by including other forms of local movement between populations
\citep{Viboud:2006p50,Balcan:2009p8522}. This work has again focused
on the spread of influenza, mixing long-distance air travel with
shorter range commuter movements; with the model predictions by
\citet{Viboud:2006p50} showing good agreement with the observed
patterns of seasonal influenza.  An alternative form of movement
network has been inferred from the ``Where's George'' study of the
circulation of dollar bills in the USA \citep{Brockmann:2006p8431};
this provided far more information about short-range movements, but
again did not really inform about the interaction of individuals.\\

A wide variety (and in practice the vast majority) of movements are
not made by aircraft, but are regular commuter movements to and from
work. The network of such movements has also been studied in some
detail for both the UK and USA \citep{Viboud:2006p50, Hall:2007p3995,
  Danon:2009p8463}. The approaches adopted parallel the work done
using the network of passenger aircraft, but operate at a much smaller
scale, and again influenza and smallpox have been the considered
pathogens. As with the aircraft network certain locations act as major
hubs attracting lots of commuters every day; however, unlike the
aircraft network there is the tendency for the network to have a
strong daily signature with commuters moving to work during the day
but travelling home again in the evening \citep{Keeling:2010p8462}. As
such the commuter network can be thought of as heterogeneous,
locally-clustered, temporal and with each contact having different
strengths (according to the number of commuters making each journey);
however, to provide a complete description of population movement and
hence disease transmission requires other causes of movement to be
included \citep{Danon:2009p8463} and requires strong assumptions to be
made about individual-level interactions. The key question that can be
readily addressed from these commuter-movement models is whether a
localised outbreak can be contained within a region or whether it is
likely to spread to other nodes on the network
\citep{Hall:2007p3995}.\\

Undoubtedly one of the largest and most comprehensive data-sets of
movements between locations comes from the livestock tracing schemes
run in Great Britain, and being adopted in other European
countries. The Cattle Tracing Scheme in particular is spectacularly
detailed, containing information of the movements of all cattle
between farms in Great Britain; as such this scheme generates daily
networks of contacts between over 30,000 working farms in Great
Britain \citep{Green:2006, Robinson:2007p4945, Heath:2008,
  Vernon:2009, Brooks:2009} (Figure 1F). Similar data also exist for
the movement of batches of sheep and pigs \citep{Kiss:2006a} although
here the identity of individual animals making each movement is not
recorded. This data source has several key advantages over other
movement networks: it is dynamic, in that movements are recorded
daily; the movement of livestock is one of the major mechanisms by
which many infections are transferred between farms; and the
metapopulation assumption that cattle mix homogeneously within a farm
is highly plausible. In principle, the information in the Cattle
Tracing Scheme can be used to form an even more comprehensive network,
treating each cow as a node and creating an edge if two cows occur
within the same farm on the same day --- this would generate an
individual-level network for each day which can then be used to
simulate the spread of infection \citep{Keeling:2010p8462}.\\

The early spread of foot and mouth disease (FMD) in 2001 was primarily
due to livestock movements, particularly of
sheep~\citep{Gibbens:2001}. Motivated by this epidemic,
\citet{Kiss:2006a} conducted short simulated outbreaks of FMD on both
the sheep movement network based on 4 weeks' movements starting on 8
September 2004, and simulated synthetic networks with the same
degree distribution. Due to the short time-scales
considered (the aim being to model spread of FMD before it had been
detected), nodes were susceptible, exposed or infected but never
recovered, and network connections remained static. Simulated
epidemics were smaller on the sheep movement network than the random
networks, most likely due to disassortative mixing in the sheep
movement network. Similarly, \citet{Natale:2009} employed a static
network simulation of Italian cattle farms. Here farms were not merely
represented as nodes, but a deterministic SI system of ODEs was used
to model infection on each node essentially generating a
metapopulation model. The only stochastic part of the model was the
number of infectious individuals moved between connected farms in each
time step. This simulation model highlighted the impact of the
centrality of seed nodes (measured in several different ways) upon the
subsequent epidemics' course.

The use of static networks to model the very dynamic movement of
livestock is questionable. Expanding on earlier work,
\citet{Green:2006} simulated the early spread of FMD through movement
of cattle, sheep, and pigs. Here the livestock network was treated
dynamically, with infection only able to propagate along edges on the
day when that edge occurred; additional to this network spread, local
transmission could also occur. These simulations enabled regional
patterns of risk to a new FMD incursion to be assessed, as well as
identifying markets as suitable targets for enhanced
surveillance. \citet{Vernon:2009} considered the relationship between
epidemics predicted from dynamic cattle networks and their static
counterparts in more detail. They compared different network
representations of cattle movement in the UK in 2004, simulating
epidemics across a range of infectivity and infectious period
parameters on the different network representations. They concluded
that network representations other than the fully dynamic one (where
the movement network changes every day) fail to reproduce the dynamics
of simulated epidemics on the fully dynamic network.\\

\subsection{Contact Tracing Networks}

Contact tracing and hence the networks generated by this method can
take two distinct forms. The first is when contact-tracing is used to
initiate pro-active control. This is often the case for STIs where
identified cases are asked about their recent sexual partners, and
these individuals are traced and tested; if found to be infected, then
contact tracing is repeated for these secondary cases. Such a process
is related to the snowball sampling that was discussed earlier, with
the notable exception that tracing is only performed from known
cases. Similar contact-tracing may operate for the early stages of an
airborne epidemic (as was seen for the 2009 H1N1 pandemic), but here
the tracing is not generally iterative as contacts are generally
traced and treated so rapidly that they are unlikely to have generated
secondary cases. An alternative form of contact-tracing is when a
transmission pathway is sought between all identified cases
\citep{Klovdahl:1985p1977, Haydon:2003p8523, Riley:2003p80}. This form
of contact tracing is likely to become of ever-increasing importance
in the future when improved molecular techniques and statistical
inference allow infection trees to be determined from genetic
differences between samples of the infecting pathogen
\citep{Cottam:2008p8548}.\\

These forms of network have two main advantages, but one major
disadvantage.  The network is often accompanied by test results for
the individuals within the network, as such we not only have
information on the contact process but also on the resultant
transmission of infection. In addition, when contact tracing is only
performed to define an infection tree, there is the added advantage
that the infection process itself defines the network of contacts and
hence there is no need for human interpretation of which forms of
contact may be relevant. Unfortunately, the reliance on the infection
process to drive the tracing means that the network only reflects one
realisation of the epidemic process and therefore may ignore contacts
that are of potential importance and would be needed if the epidemic
was to be simulated; therefore while they can inform about past
outbreaks they have little predictive power.\\

\subsection{Surrogate Networks}

Obtaining large-scale and reliable information on who contacts whom is
obviously very difficult, therefore there is a temptation to rely on
alternative data sets where network information can be extracted far
more easily and where the data is already collected. As such the
movement networks and contact tracing networks discussed above are
examples of such surrogate networks, although their connection to the
physical processes of infection transmission are far more clear. Other
examples of networks abound \citep{Liljeros:2001p5552, NRev, Boccaletti:2006p52,
  Newman:2006}; and while these are not directly relevant for the
spread of infection they do provide insights into how networks form
and grow --- structures that are commonly seen in surrogate networks
are likely to arise in the types of network associated with disease
transmission. One source of network information that would be
fantastically rich, and also highly informative (if not immediately
relevant) is the network of friendships and contacts on social
networking sites (such as Facebook); some sites have made data on
their social networks available, and these data have been used to
examine a range of sociological questions about online
interactions~\citep{Boyd:2007}.

\subsection{Theoretical Constructs}  \label{ThNet}

Given the huge complexity involved in obtaining large scale and
reliable data on real transmission networks many researchers have
instead relied on theoretically constructed networks. These networks
are usually highly simplified but aim to capture some of the known (or
postulated) features of real transmission networks --- often the
simplifications are so extreme that some analytical traction can be
gained. Here we briefly outline some of the commonly used theoretical
networks and identify which features they capture; some of the results
of how infection spreads on such networks are discussed more fully in
section \ref{subsecanalytic}.

\subsubsection{Configuration Networks}

One of the simplest forms of network is to allow each individual to
have a set of contacts that it wishes to make (in more formal language
each node has a set of half-links), these contacts are then made at
random with other individuals based on the number of contacts that
they wish to make (half-links are randomly connected)
\citep{Molloy:1998}. This obviously creates a network of contacts (Figure 1C).
The advantage of these configuration networks is that because they are
formed from many randomly connected individuals there are no short
loops within the network and a range of theoretical results can be
proved ranging from conditions for invasion \citep{Fisher:1961,
  Nickel:1983, Molloy:1995} to descriptions of the temporal dynamics
\citep{Ball:2008}. Unfortunately, the elements that make these
networks amenable to theoretical analysis --- the lack of
assortativity, short loops or clustering --- are precisely factors
that are thought to be important features of real networks.\\

An alternative formulation that offers a compromise between
tractability and realism occurs when individuals that exist in fully
interconnected cliques have randomly assigned links within the entire
population \citep{Ball:2008, House:2008} (Figure 1D). As such these networks mimic
the strong interactions within families and the weaker contacts
between them. While such models offer a significant improvement over
configuration networks, and capture the known importance of the
household in transmission, they make no allowance for clustering
between households due to spatial proximity. Hierarchical metapopulation models \citep{Watts:2005}
allow for this form of additional structure, where households (or other groupings) are themselves 
grouped in an ascending hierarchy of clustering.  \\

\subsubsection{Lattices and Small Worlds}

Both lattice networks and small world networks begin with the same
formulation: individuals are regularly spaced on a grid (usually in
just one or two dimensions), and each individual is connected to their
$k$ nearest neighbours --- these connections define a lattice. The
advantage of such networks is that they retain many elements of the
initial spatial arrangement of points, and hence contain both many
short loops as well as the property that infection tends to spread
locally. There is a clear link between such lattice-based networks and
the field of probabilistic cellular automata
\citep{LEBOWITZ:1990p8649, Rhodes:1997}. The fundamental difficulty
with such lattice models is that the presence of short loops and
localised spread mean that is it difficult (if not impossible) to
prove exact results and hence large-scale multiple simulations are
required.\\

Small world networks improve upon the rigid structure of the lattice
by allowing a low number of random contacts across the entire space
(Figure 1E). Such long range contacts allow infection to spread
rapidly though the population and vastly reduce the shortest
path-length between individuals \citep{Watts:1998p4586} --- this is
popularly known as six degrees of separation from the concept that any
two individuals on the planet are linked through at most six friends
or contacts \citep{Travers:1969}. Therefore small world networks offer
a step towards reality, capturing the local nature of transmission and
the potential for long-range contacts \citep{Boots:1999, Boots:2004},
however they suffer from neglecting heterogeneity in the number of
contacts and the tight clustering of contacts within households or
social settings.\\

\subsubsection{Spatial Networks}

Spatial networks, as the name suggests, are generated using the
spatial location of all individuals in the population; as such
lattices and small worlds are a particular form of spatial
network. The general methodology initially positions each individual
$i$ at a specific location ${\underline x}_i$, usually these locations
are chosen at random but clustered spatial distributions have also
been used \citep{Badham:2008p8668}. Two individuals (say $i$ and $j$)
are then probabilistically connected based upon the distance between
them; the probability is given by a connection kernel which usually
decays with distance such that connections are predominantly
localised. These spatial networks (especially when the underlying
distribution of points is clustered) have many features that we expect
from disease networks, although it is unclear if such simple
formulations can be truly representative.\\

\subsubsection{Exponential Random Graphs}

In recent years, there has been growing interest in exponential random
graph models (ERGMs) for social networks, also called the p* class of
models. ERGMs were first introduced in the early 1980's by
\cite{Holland1981} based on the work of \cite{Besag1974}. More
recently Frank and Strauss studied a subset of those, that have the
simple property that the probability of connection between two nodes
is independent of the connection between any other pair of distinct
nodes. \citep{Frank1986}.  This allows the likelihood of any nodes
being connected to be calculated conditional on the graph having
certain network properties. Techniques such as Markov Chain Monte
Carlo can then be used to create a range of plausible networks that
agree with a wide variety of information collected on network
structures even if the complete network is unknown
\citep{Handcock2004,Robins2004}. Due to their simplicity, ERGMs are
widely used by statisticians and social network analysts
\citep{Robins2007}.  Despite significant advances in recent years
(e.g. \cite{Goodreau2007}), ERGMs still suffer from problems of
degeneracy and computational intractability for large network sizes,
which has limited their use in epidemic modelling.

\subsection{Expected Network Properties}

Here we have shown that a wide variety of network structures have been
measured or synthesised to understand the spread of infectious
diseases. Clearly, with such a range of networks no clear consensus
can be drawn on the types of underlying network structures that are
generally present; in part this is because different studies have
focused on different infectious diseases and different diseases
require different transmission routes. However, three factors emerge
that are key components of epidemiological networks: heterogeneity in
the number of contacts such that some individuals are at a higher risk
of both catching and transmitting infection; clustering of contacts
such that groups of individuals are often highly interconnected; and
some reflection of spatial separation such that contacts usually form
locally, but occasional long-range connections do occur. \\

Three fundamental problems still exist in the study of
networks. Firstly, are there relatively low-dimensional ways of
capturing key aspects of a network's structure? What constitutes a key
aspect will vary with the problem being studied, but for
epidemiological applications it should be hoped that a universal set
of network characteristics may emerge. There is then the task of
assessing reasonable and realistic ranges for these key variables
based on values computed for known transmission networks ---
unfortunately very few transmission networks have been recorded in any
degree of detail, although modern electronic devices may simplify the
process in the future. Secondly, there is the related statistical
problem of inferring plausible complete networks from the partial
information collected by methods such as contact tracing. This is
equivalent to seeking an underlying model for the network connections
that is consistent with the known partial information, and hence has
strong resonance with the more mechanistically motivated models in
section \ref{IEN_subsect}. Even when the network is fully realised
(and an epidemic observed) there is considerable statistical
difficulty in attributing risk to particular contact types.  Finally,
there are the key questions of predicting the dynamics of infection on
any given network --- and while for many complex networks direct
simulation is the only approach, for other simplified networks some
analytical traction can be achieved, which helps to provide more
generic insights into which elements of network structure are most
important. These three key areas are discussed below.\\

\section{Network Properties}
Real networks can exhibit staggering levels of complexity. The challenge faced by researchers is to try and make sense of these structures and reduce the complexity in a meaningful way. In order to make any sense of the complexities present, researchers  over several decades have defined a large variety of measurable properties that can be used to characterise certain key aspects \citep{Albert:2002p4866, NRev, Newman:2006}. Here we describe the definitions of the most important characterisations of complex networks (in our view), and outline their impact on disease transmission models.

\subsection{Components}
In general, networks are not necessarily connected; in other words, all parts of the network are not reachable from all others. The component to which a node belongs is that set of nodes that can be reached from it by paths running along edges of the network. A network is said to have a {\em giant} component if a single component contains the majority of nodes in the network. In directed networks (one in which each edge has an associated direction) a node has both an in-component from which the node can be reached, and an out-component can be reached from that node. A strongly connected component (SCC) is the set of nodes in the network in which each node is reachable from every other node in the component. \\

The concept of a giant component is central when considering disease propagation in networks. The extent of the epidemic is necessarily limited to the number of nodes in the component that it begins in, since there are no paths to nodes in other components. In directed networks, in the case of a single initial infected individual, only the out-component of that node is at risk from infection. More generally, the strongly connected component contains those nodes that can be reached from each other. Members of the strongly connected component are most at risk from infection imported at a random node, since a single introduction of infection will be able to reach all nodes in the component. 

\subsection{Degrees, Distributions and Correlations}

The {\em degree} is defined as the {\em number of neighbours} that a node has and is most often denoted as $k$. In directed graphs, the degree has two components, the number of incoming edges $k^{in}$, (in-degree), and the number of outgoing edges $k^{out}$, (out-degree). 
The degree distribution is defined as the set of probabilities, $P(k)$, that a node chosen at random will have degree $k$.   Plotting the distribution of degrees of nodes is one of the most basic and important ways of characterising a given network (Figure 2). In addition, useful characterisations are obtained by calculating the moments of the degree distribution. The $n^{th}$ moment of $P(k)$ is defined as:
\[
\langle k^n\rangle =\sum_k k^nP(k),
\]
\noindent with the first moment, $\langle k\rangle$, being the average degree, the second, $\langle k^2\rangle$ allowing us to calculate the variance $\langle k^2\rangle - \langle k\rangle^2$, and so on. \\

The degree distribution is one of the most important ways of characterising a network as it naturally captures the heterogeneity in individuals' potential to become infected as well as cause further infection. Intuitively, the higher the number of edges a node has, the more likely it is to be a neighbour of an already infected node. Also, the more neighbours a node has, the more likely it is to cause a large number of onward cases. Thus, knowing the form of $P(k)$ is crucial for the understanding of the spread of disease. In random networks of the type studied by Erd\"{o}s and R\'{e}nyi, $P(k)$ follows a binomial distribution, which is effectively Poisson in the case of large networks. Most real social networks have distributions that are significantly different from the random case. \\

For the extreme case of $P(k)$ following an unbounded power law and assuming equal transmission across all edges, \citet{pastor01b} showed that the classic result of the epidemic threshold from mean field theory \citep{AndersonMay} breaks down. In real transmission networks, the distribution of degree is often heavily skewed, and occasionally follows a power law \citep{Liljeros:2001p5552}, but is always bounded, leading to the recovery of epidemic threshold, but one which is much lower than expected in evenly mixed populations  \citep{Lloyd:2001p6737}.\\
\begin{figure}[h!]
\centerline{\includegraphics[width=\columnwidth]{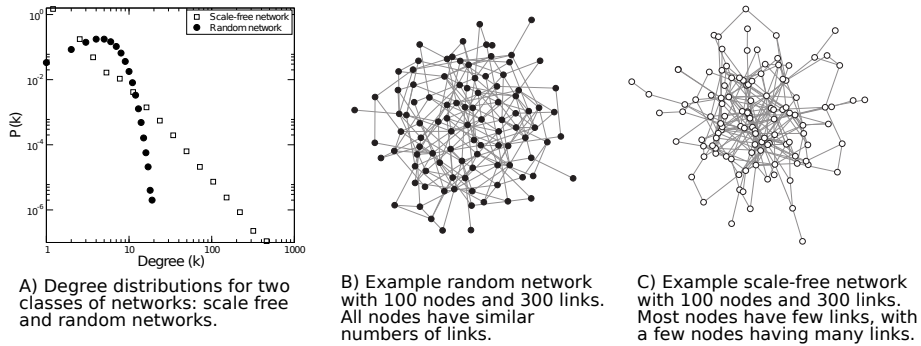}}
\caption{Comparison of random and scale-free networks}
\end{figure}

The degree distribution provides very useful information on uncorrelated networks such as those produced by configuration models. However, real networks are in general correlated with respect to degree; that is, the probability of finding a node with given degree, $k$, is dependent on the degree of the neighbours of that node, $k'$, which is captured by the conditional probability $P(k'\mid k)$. To characterise this behaviour several measurements have been proposed. The most straightforward, and probably most useful measure is to consider the average degree of the neighbours of a node: 

\[
k_{nn,i}=\frac{1}{k_i}\sum_{j \in \textrm{Nbrs}_i} k_j
\]

\noindent where the sum of degrees is made over the neighbours (Nbrs) of $i$. One can then calculate the average of $k_{nn}$ over all nodes with degree $k$ which is a direct measure of the conditional probability $P(k'\mid k)$, since
\[
k_{nn}(k)=\sum_{k'} k'P(k'\mid k).
\]

\noindent When $k_{nn}(k)$ increases with $k$, the network is said to be assortative on the degree, that is, high-degree nodes have a tendency to link to other high degree nodes, a behaviour often observed in social networks.  Other types of networks, such as the internet at router level, show the converse behaviour, i.e., nodes of high degree tend to link to nodes with low degree \citep{Newman02,NRev}.\\

Characterising degree correlations is important  for understanding disease spread. The classic example is  the existence of strong correlations in sexual networks which were shown to be a key factor in understanding HIV spread \citep{Gupta:1989p4513}. More recently, mean field solutions of the SIS model on networks have shown that both the speed and extent of an epidemic are dependent on the correlation pattern of the substrate network \citep{Bogunya02,Eguiluz:2002p10330}. 

\subsection{Distances} 

In a network, the {\em shortest path} between two nodes $i$ and $j$, is the path requiring the smallest number of steps to reach $j$ from $i$, following edges in the network. There may be (and often there is) more than one shortest path between a pair of nodes. The distance between any pair of nodes $d_{i,j}$ is the minimal number of steps required to reach $j$ from $i$, that is the number of steps in the shortest path. The average distance, $\langle d\rangle$ is the mean of the distances between all pairs of nodes and measures the typical distance between nodes: 
\[
\langle d\rangle=\frac{1}{N(N+1)}\sum_{i \neq j}d_{i,j},
\]
\noindent where $N$ is the number of nodes in the network. The diameter of the network is defined as the maximum shortest path distance between a pair of nodes in the network, max($d_{i,j}$), which measures the most extreme separation of any two nodes in the network.\\

Characterising networks in terms of the number of steps needed to reach any node from any other is also important. Real networks frequently display the small-world property, that is, the vast majority of nodes are reachable in a small number of steps. This has clear implications for disease spread and its control. Percolation approaches have shown that the effects of the small world phenomenon can be profound \citep{Moore:2000p1172}. If it only takes a short number of steps to reach everyone in the population, diseases are able to spread much more rapidly. \\

The notion of shortest distance through a network can be used to quantify how central a given node is in the network. Many measures have been used \citep{Wasserman:1994p10022}, but the most relevant of these is {\it betweenness} centrality. Betweenness captures the idea that the more shortest paths pass through a node, the
more central it is in the network. So, betweenness is simply defined as the {\it proportion} of shortest paths that pass through a single node. 

\[
B_i = \frac{\mbox {\# shortest paths through } i}{N(N-1)},
\]
\noindent where $N$ is the number of nodes in the network, and the denominator quantifies the total number of shortest paths in the network. In terms of disease spread, identifying those nodes with high betweenness will be important. Central nodes are likely to become infected early on in the epidemic, and are also key targets for intervention \citep{Bell:1999}. \\

\subsection{Clustering} \label{Cluster}
An important example of an observable property of any network is the {\em clustering
coefficient}, $\phi$, a measure of the the {\em local density} of a graph.  In social 
network terms, this quantifies the likelihood that  the friend of your friend is also your friend. It is defined as
the probability that two neighbours of a node will also be neighbours of each other and can be expressed as follows:
\[
\phi=\frac{3\times \mbox{\# of triangles in the network}}{\mbox{\# of
connected triples}},
\]
\noindent where a {\em connected triple} means a single node with edges to a pair of others. $\phi$ measures the fraction of triples that also form part of a triangle. The factor of three accounts for the fact that each triangle is found in three triples and guarantees that $0\leq \phi \leq1$ (and its inclusion depends on the way that triangles in the network are counted).\\

Locally, the clustering coefficient for each node, $i$, can be defined as the fraction of triangles formed through the immediate neighbours of $i$ \citep{Watts:1998p4586}.
\[
\phi_i=\frac{\mbox{\# triangles centered on $i$}}{\mbox{\# triples centered on $i$}}.
\]
The clustering property of networks is essential to the understanding of
transmission processes. In clustered networks, rapid local depletion of
susceptible individuals plays a hugely important role in the dynamics of spread
\citep{Keeling:1999,Eames:2002}; for a more analytic treatment of
this, see section~\ref{subsecanalytic} below.

\subsection{Subgraphs}

Degree and clustering characterise some aspects of network structure at an
individual level. Considering distances between nodes provides information
about the global organisation of the network. Intermediate scales are also
present and characterising these can help in our understanding of network
structure and therefore the dynamics of spread. \\

At the simplest level, networks can be thought of being comprised of a
collection of subgraphs. The simplest subgraph, the {\it clique}, is defined as
a group of more than two nodes where all the nodes are connected to each other
by means of edges in both directions. In other words, a clique is a fully
connected subgraph, with the smallest example being a triangle. This is a
strong definition and one which is only fulfilled in a limited number of cases,
most notably households (see Figure 1D, section \ref{subsecanalytic} and
\citet{House:2008}).  {\it n-cliques} relax the above constraint, while
retaining its basic premise. The shortest path between all the nodes in a
clique is one. Allowing this distance to take higher values, one arrives at the
definition of {\it n-cliques}, which are defined as a subgroups of the graph
containing more than two nodes where the maximum shortest path distance between
any two nodes in the group is $n$. Over the years many variants of these basic
ideas have been formalised in the social network literature and a good summary
can be found in \citet{Wasserman:1994p10022}.\\

Considering higher order structures can be very informative but is more
involved. Milo and co-workers began by looking for specific patterns of
connections between nodes in small sub-graphs, dubbed {\it motifs}. Given a
connected sub-graph of size 3 (for example), there are 13 possible motifs.
Statistically, some of these appear more often and are found to be
over-represented in certain real networks compared to random networks
\citep{Milo:2002p1036}. Understanding the motif composition of a complex
network has been shown to improve the predictive power of deterministic models
of transmission when motifs are explicitly modelled  (see section
\ref{subsecanalytic} and \citet{House:2009p9133}).\\

In the above definitions, a subgraph has been defined only in reference to
itself. A different approach is to compare the number of internal edges to the
number of external edges, arising from the intuitive notion that a {\it
community} will be denser in terms of edges than its surroundings. One such
definition, the definition of community in the {\it strong} sense, is defined
as a subgraph in which each node has more edges to other nodes within the
subgraph than to any other nodes in the network. Again, this  definition is
quite restrictive, and in order to relax these constraints, the most commonly
used (and most intuitive) definition of communities is groups of nodes that
have a high density of edges within them and a lower density of edges between
groups. This intuitive definition is behind the most widely used approach for
studying community structure in networks. Newman and Girvan formalised this in
terms of the {\it modularity} measure $Q$ \citep{NG}. Given a particular
network which is partitioned into communities, the modularity measure compares
the expected number of edges within communities to the actual number of edges
within communities. 

Although the impact of communities in transmission processes has not been fully
explored, a few studies have shown it can have a profound impact on disease
dynamics \citep{Buckee:2007p926,Salathe:2010p9807}. An alternative measure of
how ``well-knit" a graph is, named conductance \citep{Kannan2004}, most widely
used in the computer science literature has also been found to be important in
a range of networks \citep{Leskovec2008a}.

\subsection{Higher Dimensional Networks} 

All of the above definitions have concentrated on networks where the edges
remain unchanged over time and all edges have equal weight. Both of these
constraints can naturally be relaxed, but generally this calls for a
higher-dimensional characterisation of the edges within the network.  It is a
matter of common experience that social interactions which can lead to
infection do change, with some contacts being repeated regularly, while others
are more sporadic. The frequency, intensity and duration of contacts are all
time-varying. How these inherently dynamic networks are represented for the
purposes of modelling can have a significant impact on the model outcomes
\citep{Vernon:2009,Kao:2006p527}. However, capturing the structure of such
dynamic networks in a parsimonious manner remains a substantial challenge. More
work has been done on weighted networks, as these are a more straight-forward
extension of the classical presence-absence networks \citep{Barrat:2004,
Newman:2004}. In terms of disease spread, the movement networks discussed in
section \ref{Movements} are often considered as weighted
\citep{Hufnagel:2004p4303, Viboud:2006p50, Robinson:2007p4945}.\\

In the sections that follow we discuss how these network properties
can be inferred statistically and the improvements in our
understanding of the transmission of infection in networks that have
come as a result. \\

\section{Model Formulation}
\subsection{Techniques for Simulation}

One of the key advantages of the simulation of disease processes on
networks is that it enables the study of systems that are too complex
for analytical approaches to be tractable. With that in mind, it is
worth briefly considering efficient approaches to disease simulation
on networks.\\

There are two main types of simulation model for infectious diseases
on networks: discrete-time and continuous-time models; of these,
discrete-time simulations are more common, so we discuss them
first. In a discrete-time simulation, at every time-step disease may
be transmitted along every edge from an infectious node to a
susceptible node with a particular probability (which may be the same
for all extant edges, or may vary according to properties of the two
nodes or the edge). Also, nodes may recover (becoming immune, or
reverting to being susceptible) during each time-step. Within a
time-step, every infection and recovery event is assumed to occur
simultaneously. In a dynamic network simulation, the network is
typically updated every time-step --- for example, in a livestock
movement network, during time-step $x$, infection could only transmit
down edges that occurred during time-step $x$. Clearly, in a directed
network, infection may only transmit in the direction of an edge.\\

Whilst algorithms for discrete-time simulations are not complex, some
simple implementation techniques (arising from the observation that
most networks of epidemiological interest are sparse) can
significantly enhance software performance. In a directed network with
$N$ nodes, there are $N(N-1)$ possible edges; in a sparse network with
mean node degree $k$, there are $Nk\ll N(N-1)$ edges. Accordingly,
rather than representing the network as an $N$ by $N$ array, where the
element in each array is 0 if the edge is absent, nonzero otherwise,
it is usually more efficient to maintain a list of the neighbours of
each node. Then, if a list of infected nodes is maintained during a
simulation run, it is straightforward to consider each susceptible
neighbour of an infected node in turn and test if infection is
transmitted to that node. Additionally, a fast high-quality
pseudo-random number generator such as the Mersenne Twister should be
used \citep{Matsumoto:1998}. The ``contagion'' software package
implements these techniques (amongst others), and is freely
available~\citep{Contagion}. \\

The alternative approach to simulating disease processes on networks
is to simulate a series of stochastic Markovian events --- the
continuous-time approach.  Essentially, given the state of the system,
it is possible to calculate the probability distributions of when
possible subsequent events (i.e. recovery of an infectious node or
infection of a susceptible node) will occur. Random draws from these
distributions are then made to determine which event occurs next, the
state of the system updated, and the process repeated. This approach
was pioneered by Gillespie to study the dynamics of chemical
reactions~\citep{Gillespie:1977}; it is, however, computationally
intensive, so approximations have been developed. The $\tau$-leap
method~\citep{Gillespie:2001}, where multiple events are allowed to
occur during a time period $\tau$, is clearly related to the
discrete-time formulation discussed above. However, the ability to
allow $\tau$ to vary during a simulation to account for the processes
involved ~\citep{Cao:2006} has potential benefits.\\

The continuous-time approach is clearly in closer agreement with the
ideal of standard disease models, however utilising this method may be
computationally prohibitive especially when large networks are
involved. Discrete-time models may provide a viable alternative for
three main reasons. Firstly, as the time-steps involved in the
discrete-time model become sufficiently small, we would expect the two
models to converge. Secondly, inaccuracies due to the discrete-time
formulation are likely to be less substantial in network models
compared to random-mixing models, providing two events do not occur in
the same neighbourhood during the same time-step. Finally, the daily
cycle of contacts that regulate most of our lives means that using
time-steps of less than 24 hours may falsely represent the temporal
accuracy that can be attributed to any simulation of the real
world. \\

\subsection{Analytic Methods} \label{subsecanalytic}

In this section we use the word `analytic' broadly, to imply models
that are directly numerically integrable, without the use of Monte
Carlo simulation methods, rather than systems for which all results
can be written in terms of fundamental functions, of which there are
very few in epidemiology. Analytic approaches to transmission of
infection on networks fall into three broad categories. Firstly, there
are approaches that calculate exact invasion thresholds and final
sizes for special networks. Secondly, there are approaches for
calculating exact transient dynamics, including epidemic peak heights
and times, but again these only hold in special networks. Finally,
there are approaches based on moment closure that are give
approximately correct dynamics for a wide class of networks.\\

Before considering these approaches on networks, it is worth
considering what is meant by non-network mixing, and showing
explicitly how this can derive the standard transmission terms from
familiar differential equation models. Non-network mixing can be taken
to have one of two meanings: either that every individual in the
population is weakly connected to every other (the mean-field
assumption), or that an Erd\"{o}s-R\'{e}nyi random graph defines the
transmission network, depending on context. To see how this determines
the epidemic dynamics, we consider a population of $N$ individuals,
with a homogeneous independent probability $q$ that any pair of
individuals is linked on the network, which gives each individual a
mean number of edges $\bar{n} = q (N-1)$. We then assume that the
transmission rate for infection across an edge is $\tau$ and that the
proportion of the population infectious at a time $t$ is $I(t)$; then
the force of infection experienced by an average susceptible in the
population is $\bar{n} \tau I(t) \equiv \beta I(t)$.  The quantity
$\beta$ therefore defines a population-level transmission rate that
can be interpreted in one of two ways as $N\rightarrow\infty$. In the
case where the population is assumed to be fully connected, the limit
is that $q$ is held at unity, and so $\tau$ is reduced to as $N$ is
increased to hold $q(N-1)\tau$ constant. In the case where the
population is connected on a random graph, $q$ is reduced as $N$ is
increased to hold $\bar{n}$ constant.\\

In either case, having defined an appropriate population-level
transmission rate, a stochastic susceptible-infectious model of
transmission is defined through a Markov chain in which a population
with $X$ susceptible individuals and $Y$ infectious individuals
transitions stochastically to a population with $X-1$ susceptible
individuals and $Y+1$ infectious individuals at rate $\beta X Y /
(N-1)$.  Then the exact mean behaviour of such a system in the limit
$N\rightarrow \infty$ then has its transmission behaviour captured by
\begin{equation}
\dot{S} = - \beta S(t) I(t) \text{ ,}
\end{equation}
where $S,I$ are the proportion of individuals susceptible and infectious
respectively. The mathematical formalism behind deriving such sets of ordinary
differential equations from Markov chains is given by~\citet{Kurtz:1970}, and a
summary of the application of this methodology to infectious disease modelling
is given in~\citet{DiekHees00}. However, it should be clear that equation (1) is familiar as the basis of all random-mixing epidemiological models.\\

In the case of exponentially-distributed infectious periods and
recovery from infection offering long-lasting immunity, the standard
\textit{SIR} equations provide an exact description of the mean
behaviour of this system.  Nevertheless, the existence of waning
immunity, a latent period between an individual becoming infected and
being able to transmit infection, and non-exponentially distributed
recovery periods are also important for epidemiological
applications~\citep{AndersonMay,KeelingRohani,Ross:2010p7650}. These can often be
incorporated into analytical approaches through the addition of extra
disease compartments, which necessitates extra algebraic and
computational effort but typically does not require a fundamental
conceptual re-evaluation.  Sometimes significant additional complexity
does not even modify quantitative epidemiological results---for
example, regardless of the rate of waning immunity, length of latent
period, or infectious period distribution, if the mean infectious
period is $T$ then the basic reproductive ratio is
\begin{equation}
R_0 = \beta T \text{ .}
\end{equation}
The estimation of this quantity for complex disease histories, from
data likely to be available, is considered
by~\citet{Wallinga:2007}. We therefore focus on the transmission
process, since this is most affected by network structure, and other
elements of biological realism typically act at the individual
level. An important caveat to this, however, is when an infected
individual's level of transmissibility varies over the course of their
infectious period, which sets up correlations between the processes of
transmission and recovery that pose a particular challenge for
analytic work, especially in structured populations, as noted by
e.g.\ \citet{Ball:2009}.

\subsubsection{Exact Invasion}
For non-network mixing, the threshold for invasion is given by the
basic reproductive ratio $R_0$, defined as the expected number of
secondary infectious cases created by an average primary infectious
case in an otherwise wholly susceptible population. In structured
populations, this verbal definition is typically altered to be the
secondary cases caused by a typical primary case once the dynamical
system has settled into its early asymptotic behaviour. As such, the
threshold for invasion is $R_0=1$: for values above this an infection
can grow in the population and the disease can successfully invade;
for values below it each chain of infection is doomed to eventual
extinction. Values of $R_0$ can be measured directly during the course of
an epidemic by detailed contact tracing, however there are
considerable statistical issues concerning censoring and data
quality.\\

Provided there are no short closed loops in the network, $R_0$ can
be defined through a next-generation matrix:
\begin{equation}
K_{km} = \frac{[km](m-1)}{m[m]} p
\label{nextgenK}
\end{equation}
where $K_{km}$ defines the number of cases in individuals with $k$
contacts from an individual with $m$ contacts during the early stages
of the epidemic. Here and elsewhere in this section we use square
brackets to represent the numbers of different types on the network;
hence $[m]$ is the number of individuals with $m$ edges in the network
and $[km]$ is the number of edges between individuals with $k$ and $m$
contacts respectively. In addition $p$ is the probability of infection
eventually passing across the edge between a
susceptible-infectious pair (for Markovian recovery rate $\gamma$ and
transmission rate $\tau$ this is given by $p=\tau/(\tau +
\gamma)$). The basic reproductive ratio is given by the dominant
eigenvalue of the next-generation matrix:
\begin{equation}
R_0 = || (K_{km}) || \text{ .}
\label{unclustR0}
\end{equation}
This quantity corresponds to the standard verbal definition of the
basic reproductive ratio, and correspondingly the invasion threshold
is at $R_0=1$.\\

Once an appreciable number of short closed loops are present in the
network, exact threshold parameters can still sometimes be defined,
but these typically depart from the standard verbal definition of
$R_0$.  For example,~\citet{Ball:2009} consider a branching process on
cliques (households) connected to each other through
configuration-model edges --- cliques are connected to each other at
random (Figure 1D). By considering the number of secondary cliques
infected by a clique with one initial infected individual, a threshold
called $R_*$ can be defined. (For the configuration-model of
households where each household is of the same size and each
individual has the same number of random connections outside the
household, the threshold $R_*$ is given later as equation \ref{RSTAR};
however the methodology is far more general).  The calculation of the
invasion threshold for the recently defined Triangular Configuration
Model~\citep{Miller:2009, Newman:2009} involves calculating both the
expected number of secondary infectious individuals and triangles
rather than just working at the individual level. \citet{Trapman:2007}
deals with how these sort of results can be related to more general
networks through bounding. A general feature of clustered networks for
which exact thresholds have been derived so far is that there is a
local-global distinction in transmission routes, with a general theory
of this given by~\citet{Ball:2002}, where an `overlapping groups' and
`great circle' model are also analysed. Nevertheless, care still has
to be taken in which threshold parameters are mathematically well
behaved and easily calculated~\citep[e.g.][]{L.Pellis11062009}.\\

\subsubsection{Exact Final Size}

The most sophisticated and general way to obtain exact results for the expected
final size of a major outbreak on a network is called the
\textit{susceptibility set} argument and the most general version is
currently given by~\citet{Ball:2009}. We give an example of these kind of
arguments from~\citet{Diekmann:1998}, who consider the simpler case of a
network in which each individual has $n$ contacts. Where there is a probability
$p$ of infection passing across a given network link (so for transmission and
recovery at rates $\tau$ and $\gamma$ respectively, $p = \tau/(\tau+\gamma )$)
the probability that an individual avoids infection is given by
\begin{equation}
\begin{aligned}
S_{\infty} & = \left(1 - p + \tilde{S} p\right)^{n} \text{ ,}\\
\tilde{S} & = \left(1 - p + \tilde{S} p\right)^{n-1} \text{ .}
\label{FinalSize}
\end{aligned}
\end{equation}
Here, a two-step process is needed because in an unclustered, regular graph two
generations of infection are needed to stabilise the network correlations and
so the auxiliary variable $\tilde{S}$ must also be solved for. Once this and
$S_\infty$ are known, the expected attack rate is $R_\infty = 1 - S_\infty$. 

\subsubsection{Approximate Final Size}

The main way to calculate approximate final sizes is given by percolation-based
methods. These were reviewed by~\citet{Bansal:2007} and also
in~\cite{Newman:2010}. Suppose we remove a fraction $\varphi$ of links from
the network, and can derive an expression for the fraction of nodes remaining
in the giant component of the network, $f(\varphi)$. Then
\begin{equation}
R_\infty \approx f(1 - p) \text{ ,}
\end{equation}
and an invasion threshold is given by the value of $p$ for which this final
size becomes non-zero in the `thermodynamic limit' of very large network size.
This approach is not exact for clustered graphs, but for unclustered graphs
exact results like~\eqref{FinalSize} are reproduced.

\subsubsection{Exact Dynamics}
\label{exdyn}
Some of the earliest work on infectious diseases involved the exact
solution of master equations (where the probability of the population
being in each possible configuration is calculated) on small, fully
connected graphs as summarised in~\citet{Bailey:1975}. The rate at
which the complexity of the system of master equations grows means
that these equations quickly become too complex to integrate for the
most general network.  The presence of symmetries in the network,
however, does mean that automorphism-driven lumping is one way to
manipulate the master equations (whilst preserving the full stochastic
information about the system) for solution~\citep{Kiss:2010}. At
present, this technique has only been applied to relatively simple
networks, however there are no other highly general methods of
deriving exact lower-dimensional systems of equations from the master
equations.\\

Nevertheless, other specific routes do exist that allow exact systems
of equations of lower dimensionality to be derived for special
networks. For static networks constructed using the configuration
model (where individuals have heterogeneous degree but connections are
made at random such that the presence of short loops can be ignored in
a large network, see Figure 1C), an exact system of equations for
\textit{SIR} dynamics in the limit of large network size was provided
by~\citet{Ball:2008}. This construction involves attributing to each
node an `effective degree', which starts the epidemic at its actual
degree, and measures connections still available as routes of
infection and is therefore reduced by transmission and recovery. Using
notation consistent with elsewhere in this paper (and ignoring the
global infection terms that were included by Ball and co-workers) this
yields the relatively parsimonious set of equations:
\begin{equation}
\begin{aligned}
\dot{S}_k & = - \rho \left( (\tau + \gamma) k S_k
  - \gamma (k+1) S_{k+1} \right) \text{ ,}\\
\dot{I}_k & = \tau \left((k+1)I_k - kI_k \right) - \gamma I_k
  + \rho \bigg( (k+1) \left( \tau(S_{k+1} + I_{k+1} ) + \gamma I_{k+1}
  \right) - k (\tau + \gamma) I_k \bigg) \text{ ,}\\
\rho & := \frac{\sum_k k I_k}{\sum_l l (S_l + I_l )} \text{ .}
\label{exactCM}
\end{aligned}
\end{equation}
Here $S_k, I_k$ are the proportion of effective degree $k$ susceptible
and infectious individuals respectively. Hence for a
configuration-network where the maximum degree is $K$ we require just
$2K$ equations to retrieve the exact dynamics.\\

While $R_0$ can be derived using expressions like~\eqref{unclustR0},
calculation of the asymptotic early growth rate $r$ requires systems of ODEs
like \eqref{exactCM}. If we assume 
that transmission and recovery are Markovian processes with rates $\tau$ and
$\gamma$ respectively, two measures of early behaviour are
\begin{align}
R_0^{\mathrm{CM}} & = \frac{\left<n(n-1)\right>}{\left<n\right>}
 \frac{\tau}{\tau + \gamma} \text{ ,} &
r^{\mathrm{CM}} & = \frac{\left<n(n-2)\right>}{\left<n\right>}\tau 
 - \gamma \text{ ,} 
\end{align}
where $<.>$ informs about the average over the degree distribution. 
These quantities tell us that the susceptibility to invasion of a network increases with both the mean
and the variance of the degree distribution. This closely echos the
results for risk-structured models \citep{AndersonMay} but with an extra term of $-1$
due to the network, 
representing the fact that the route through which an individual
acquired infection is closed off for future transmission events.\\

For more structured networks with a local-global distinction, there are two
limits in which exact dynamics can also be derived.  If the network is composed
of $m$ communities of size $n_1, \ldots, n_m$, with the between-community (global) mixing determined by
a Poisson process with rate $\bar{n}_G$ and the within-community (local) mixing
determined by a Poisson process with rate $\bar{n}_L$, then in the limit as the
communities become large, $n_i\rightarrow\infty$, the epidemic dynamics
on the system are
\begin{equation}
\begin{aligned}
\dot{S}_a & = - S_a \left( \beta_L I_a + \alpha \sum_{b \neq a} I_b \right)
 \text{ ,}\\
\dot{I}_a & = S_a \left( \beta_L I_a + \alpha \sum_{b \neq a} I_b \right)  - \gamma I_a \text{ ,}
\end{aligned} \label{metapop}
\end{equation}
where $S_a$ and $I_a$ are the proportion of individuals susceptible and infectious in community $a$, and
\begin{equation}
\alpha = \frac{\bar{n}_G \tau}{(m-1)} \text{ ,} \qquad
\beta_L = \bar{n}_L \tau \text{ .}
\end{equation}
Hence, we have a classic metapopulation model \citep{Hanski_book_2004}, defined
in terms of Poisson local and global connections and large local community
sizes.\\

In the limit where $\bar{n}_L \rightarrow (n-1)$ and $m\rightarrow\infty$ ---
such that there are infinitely many communities of equal size and each
community forms a fully interconnected clique --- then `self-consistent'
equations such as in~\citet{Ghoshal:2004} and~\citet{House:2008} are exact.
These equations evolve the proportion of cliques with $x$ susceptibles and $y$
infecteds, $P_{x,y}$,  as well as the proportion of infecteds in the
population, $I$, as
follows:
\begin{equation}
\begin{aligned}
 I & = \frac{1}{n} \sum_{x,y} y P_{x,y} \text{ ,}\\
\dot{P}_{x,y}  
& = \gamma \left( -y P_{x,y} + (y + 1) P_{x,y+1} \right) \\
& + \tau \left( -x y P_{x,y} + (x + 1) (y - 1) 
  P_{x+1,y-1} \right) \\
& + \beta_G I \left( -x P_{x,y} + (x + 1)
  P_{x+1,y-1} \right) \text{ ,}
\end{aligned} \label{SmallCliques}
\end{equation}
where $\beta_G = \bar{n}_G \tau$. \\

Both of these two local-global models, the metapopulation model (\ref{metapop}) and the small cliques model (\ref{SmallCliques}), are
reasonably numerically tractable for modern computational resources, provided
the relevant finite number ($m$ or $n$ respectively) is not too large.  
The basic reproduction number for the first system is clearly
\begin{equation}
R_0 = \frac{1}{\gamma} (\beta_L + (m-1)\alpha) \text{ ,}
\end{equation}
while for the second, household model, invasion is determined by
\begin{equation}
R_* = \bar{n}_G \frac{\tau}{\tau +\gamma} Z^n_\infty(\tau,\gamma) \text{ ,}
\label{RSTAR}
\end{equation}
where $Z^n_\infty(\tau,\gamma)$ is the expected final size of an epidemic in a
household of size $n$ with one initial infected. 
Of course, the within- and between-community mixing for real networks is likely to
be much more complex than may be captured by a Poisson process, but these two extremes can provide
useful insights.
These models show that network structure of the form of communities reduces the potential for an infectious disease to spread, and hence greater transmission rates are required for the disease to exceed the invasion threshold.\\

\subsubsection{Approximate Dynamics}

While all the exact results above are an important guide to intuition they only
hold for very specialised networks. A large class of models exists that form a
bridge between `mean-field' models and simulation by using spatial or network
moment closure equations. These are highly versatile models. In general,
invasion thresholds and final sizes can be calculated rigorously, but exact
calculation of transient dynamics is only possible for very special networks.
If one wants to calculate transient effects in general network models---most
importantly, peak heights and times---then moment closure is really the only
versatile way of calculating desired quantities without relying on full
numerical simulation.

It is also worth noting that there are many results derived through these
`approximate' approaches that are the same as exact results, or are numerically
indistinguishable from exact results and simulation. We give some examples
below, and also note that the dynamical PGF approach~\citep{Volz:2008} is
numerically indistinguishable from the exact model~\eqref{exactCM} above for
certain parameter values~\citep{Lindquist:2010}. What is currently lacking is a
rigorous mathematical proof of exactness for ODE models other than those
outlined in section \ref{exdyn} above. While for many practical purposes the
absence of such a proof will not matter, we preserve here the conceptual
distinction between results that are provably exact, and those that are
numerically exact in all cases tested so far.\\

The idea of moment closure is to start with an exact but unclosed set of
equations for the time evolution of different units of structure. Here we show
how these can be derived by considering the rates of change of both types 
of individual and types of connected pair. Such pair-wise moment closure model are a 
natural extension to the standard (random-mixing) models, given that infection
is passed between pairs of infected individuals:
\begin{equation}
\begin{aligned}
\dot{[S_{\kappa}]} & = - \tau [S_{\kappa}\leftarrow I] \text{ ,} \\
\dot{[I_{\kappa}]} & = \tau [S_{\kappa}\leftarrow I] 
 - \gamma[I_{\kappa}] \text{ ,} \\
\dot{[S_{\kappa}S_{\lambda}]} & = - \tau [S_{\kappa}S_{\lambda}\leftarrow I] 
 + [S_{\lambda}S_{\kappa}\leftarrow I] \text{ ,} \\
\dot{[S_{\kappa}I_{\lambda}]} & = 
  \tau \left( [S_{\kappa}S_{\lambda}\leftarrow I] 
  - [I\rightarrow S_{\kappa}I_{\lambda}] 
  - [S_{\kappa}\leftarrow I_{\lambda}] \right)
  -\gamma [S_{\kappa}I_{\lambda}] \text{ ,} \\
\dot{[I_{\kappa}I_{\lambda}]} & = 
  \tau \left( [I\rightarrow S_{\kappa}I_{\lambda}] 
  + [I\rightarrow S_{\lambda}I_{\kappa}] 
  + [S_{\kappa} \leftarrow I_{\lambda}] 
  + [S_{\lambda}\leftarrow I_{\kappa}]\right) 
  -2 \gamma [I_{\kappa}I_{\lambda}] \text{ ,} \\
\dot{[S_{\kappa}R_{\lambda}]} & = -\tau [I\rightarrow S_{\kappa}R_{\lambda}] 
  + \gamma [S_{\kappa}I_{\lambda}] \text{ ,} \\
\dot{[I_{\kappa}R_{\lambda}]} & = 
  \tau [I\rightarrow S_{\kappa}R_{\lambda}] 
  + \gamma ( [I_{\kappa}I_{\lambda}] 
  - [I_{\kappa}R_{\lambda}]) \text{ .}
\label{fullpw}
\end{aligned}
\end{equation}
Here we use square brackets to represent the prevalence of different species
within the network. We also use some non-standard notation to
present several diverse approaches in a unified framework: generalised indices
$\kappa, \lambda$ represent any property of a node (such as its degree); while arrows represent the
direction of infection (and so for a directed network, the necessity that an
edge in the appropriate direction be present).

Clearly, the system~\eqref{fullpw} is not closed as it relies on the number of
connected triples, and so some form of approximate closure must be introduced
to relate the triples to pairs and nodes, which will depend on underlying
properties of the network.  Most commonly, these closure assumptions deal with
heterogeneity in node degree, assortativity, and clustering at the level of
triangles. Examples include~\citet{Keeling:1999} and \citet{Eames:2002}, where
the generalised variables $\kappa,\lambda$ above stand for node degrees
($k,l$), the triple closure is symmetric with respect to the direction of
infection, and the network is assumed to be static and non-directed. A
general way to write the closure assumption is:
\begin{equation}
[A_kB_lC_m] \approx \frac{(l-1)}{l} \left( (1-\phi ) 
   \frac{[A_kB_l][B_lC_m]}{[B_l]} 
 + \phi \frac{\bar{n} N}{km} \frac{[A_kB_l][B_lC_m][C_mA_k]}{[A_k][B_l][C_m]}
 \right) \text{ .}
 \label{fulltripleclose}
\end{equation}
where $\bar{n} \equiv \left<n\right>$ is again the average degree distribution,
and $\phi$ measures the ratio of triangles to triples as a means of capturing
clustering within the network (see section \ref{Cluster}).  The typical way to
analyse the closed system is direct numerical integration, however some
analytic traction can be gained. One example is the use of a linearising Ansatz
to derive the early asymptotic behaviour of the dynamical system.
Interestingly, when this is done for $\phi=0$ (such that there are no
triangular loops in the network) as in~\citet{Eames:2002}, the result for the
early asymptotic growth rate agrees with the exact result of
equation~\eqref{nextgenK}. In~\cite{Keeling:1999}, the differential equations
for an $n$-regular graph were also manipulated to give an expression for final
size that agreed with the exact result~\eqref{FinalSize}\\

Equation~\eqref{fulltripleclose} however, is not the only possible network
moment closure regime: \citet{Boots:2002} and \citet{Bauch:2005}
considered regimes in which closure depended on the disease state (i.e.\
triples composed of different arrangements of susceptibles and infecteds close
differently) to deal with spatial lattice-based systems and early disease invasion
respectively. For example \citet{Boots:2002} use a closure where
\begin{equation}
\begin{aligned}
\, [SOO] & \approx \varepsilon \frac{\bar{n}-1}{\bar{n}} \frac{[SO][OO]}{[O]} 
\text{ ,}\\
[IOO] & \approx \varepsilon \frac{\bar{n}-1}{\bar{n}} \frac{[IO][OO]}{[O]} 
\text{ ,}\\
[SOS] & \approx [OS] \frac{\bar{n}-1}{\bar{n}} \left(1 - \varepsilon \frac{[OO]}{[O]} - \frac{[IO]}{[O]}\right) 
\text{ ,}\\
[ABC] & \approx \frac{\bar{n}-1}{\bar{n}} \frac{[AB][BC]}{[B]}
\quad\text{for all other triples,}\\
\end{aligned}
\end{equation}
where $O$ represents empty sites within the network that are not currently
occupied by individuals, and the parameter $\varepsilon = 0.8093$ accounts for
the clustering within lattice-based networks.  \citet{House:2010a} considered a
model of infection transmission and contact tracing on a network, where the
closure scheme for $[ABC]$ triples was asymmetric in $A$ and $C$ -- this
allowed the natural conservation of quantities in a highly clustered system.\\

The work on dynamical PGF models \citep{Volz:2008} can be seen as an elegant
simplification of this pairwise approach that is valid for SIR-type infection
dynamics on configuration model networks. The equations can be reformulated as:
\begin{equation}
\begin{aligned}
S & = g(\theta) \\
\dot{ I } & = \tau p_I \theta g'(\theta) - \gamma I \\
\dot{ p_I } & = \tau p_S p_I \theta \frac{g''(\theta)}{g'(\theta)} - \tau p_I (1-p_I) - \gamma p_I \\
\dot{ p_S } & = \tau p_S p_I \left( 1 - \theta
\frac{g''(\theta)}{g'(\theta)}\right) \\
\dot{ \theta} & = - \tau p_I \theta
\end{aligned}
\end{equation}
where $g$ is the probability generating function for the degree distribution,
$p_S$ and $p_I$ correspond to the number of contacts of a susceptible that are
susceptible or infected respectively, and $\theta$ is defined as probability
that a link randomly selected from the entire network has not been associated
with the transmission of infection. Here the closure assumption is implicit in
the definition of $S$; that an individual only remain susceptible if all of its
links have not seen the transmission of infection and that the probability is
independent for each link, which is comparable to the assumptions underlying
the formulation by \citet{Ball:2008}, equation (\ref{exactCM}). The precise
link between this PGF formulation and the pairwise approach is discussed more
fully in \citet{House:2010}.\\

There are many other extensions of this general methodology that are possible.
Writing ODEs for the time evolution of triples and closing at a higher order
allows the consideration of the epidemiological consequences of varying motif
structure~\citep{House:2009}. \citet{Sharkey:2006} considered closure at
triple level on directed networks, which involved a more sophisticated
treatment of third-order clustering due to the larger repertoire of
three-motifs in directed (as compared to undirected) networks.  It is also
possible to combine stochastic and network moment
closure~\citep{Dangerfield:2008}.  Time-varying, dynamical networks,
particularly applied to sexually transmitted infections where partnerships vary
over the course of an epidemic, were considered using approximate ODE-based
models by~\citet{Eames:2004}, and~\citet{Volz:2007}.  \citet{Sharkey:2008}
considered models appropriate for local networks with large shortest path
lengths, where the generic indices $\mu,\lambda$ in~\eqref{fullpw} stand for
node numbers $i,j$ rather than node degrees $k,l$.\\

Another approach is to approximate the transmission dynamics in the standard (mean-field) differential equations models. Essentially this is a form of moment closure at the
level of pairs rather than triples. For example, in~\citet{roy:2006} the transmission rate takes
the polynomial form:
\begin{equation}
\mbox{Transmission rate to $S_k$ from $I_l$} \propto k l (S_k)^{p} (I_l)^{q} \text{ ,}
\end{equation}
where the the exponents, $p$ and $q$, are typically fitted to
simulated data but are thought to capture the spatial arrangement of susceptible and infected nodes.
Also,~\citet{Kiss:2006} suggest:
\begin{equation}
\mbox{Transmission rate to $S_k$ from $I_l$} \propto k (l-1) (S_k) (I_l) \text{ ,}
\end{equation}
as a way of accounting for each infected `losing' an edge to its infectious
parent.\\

Finally, very recent work~\citep{Volz:2010} presents a dynamical system to
capture epidemic dynamics on triangular configuration model networks; the
relationship between this and other ODE approaches is likely to be an active
topic for future work.\\

This diversity of approaches leads to some important points about methods based
on moment closure.  These methods are extremely general, and can be applied to
consider almost any aspect of network structure or disease natural history;
they can be applied to populations not currently amenable to direct simulation
due to their size; and they do not require a complete description of the
network to run---only certain statistical properties. However, there are
currently no general methods for the proposal of appropriate closure regimes, nor any
derivation of the limits on dynamical biases introduced by closure.  Therefore,
closure methods sit somewhere in between exact results for highly specialised kinds
of network and stochastic simulation where intuitive understanding and general
analysis are more difficult.\\

\subsection{Comparison of analytic models with simulation}

In the papers that introduced them, the differential-equation based approximate
dynamical systems above were compared to stochastic simulations on appropriate
networks. Two recent papers making a comparison of different dynamical systems
with simulation are~\citet{Bansal:2007} and~\citet{Lindquist:2010}. There are,
however, several issues with attempts to compare deterministic models with
simulation and also with each other.

Firstly, it is necessary to define what is meant by agreement between a smooth,
deterministic epidemic curve and the rough trajectories produced by simulation.
Limiting results about the exactness of different ODE models assume that both
the number of individuals infectious and the network size are large, and so the
early behaviour of simulations, when there are few infectious individuals, is
often dominated by stochastic effects. There are different ways to address this
issue, but even after this has been done there are two sources of deviation
of simulations from their deterministic limit. The first of these is the
number of simulations realised. If there is a summary statistic such as
the mean number of infectious individuals over time, then the confidence
interval in such a statistic can be made arbitrarily small by running
additional simulations, but agreement between the deterministic limit and a
given realisation may still be poor. The second source of deviation is the
network size. By increasing the number of nodes the prediction interval within
which the infection curve will fall can be made arbitrarily small, however
the computational resources needed to simulate extremely large networks
can quickly become overwhelming.

More generally, each approximate model is designed with a different application
in mind. Models that perform well in one context will often perform poorly in
another, and this means that `performance' of a given model in terms of
agreement with simulation will primarily be determined by the discrete network
system on which simulations are performed.

\begin{figure}[h!]
\centerline{\includegraphics[width=.9\columnwidth]{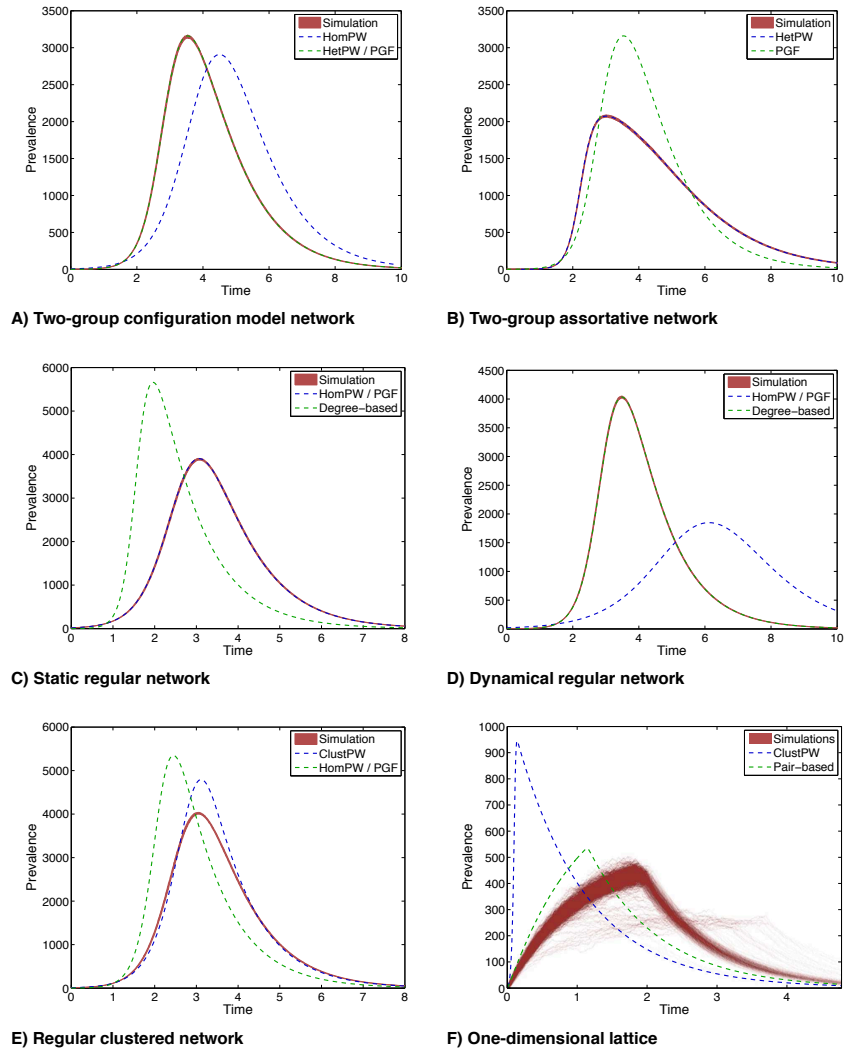}}
\caption{Comparison of simulation and deterministic models for six networks}
\label{CompareFig}
\end{figure}

The above considerations motivate the example comparisons with simulation that
we show in Figure~\ref{CompareFig}. This collection of plots is intended to
show a variety of different example networks, and the dynamical systems
intended to capture their behaviour.

In the first five plots of Figure~\ref{CompareFig}, continuous-time simulations
have their temporal origin shifted so that they agree on the time at which a
cumulative incidence of 200 is reached, and then confidence intervals in the
mean prevalence of infection are achieved through bootstrapping. The 95\%
confidence interval is shown as a red shaded region (although typically this is
sufficiently narrow it resembles a line). Six different deterministic models
are compared to simulations: HomPW is the pairwise model
of~\citet{Keeling:1999} with zero clustering; HetPW is the heterogeneous
pairwise model of~\citet{Eames:2002}; ClustPW is the improved clustered
pairwise closure of~\citet{House:2010a}; PGF is the model of~\citet{Volz:2008};
Pair-based is the model of~\citet{Sharkey:2008}, integrated using the
supplementary code from~\citet{Sharkey:2010}; and Degree-based is the model
of~\citet{pastor01b}.

Plot A shows a heterogeneous network composed of two risk groups, constructed
according to the configuration model~\citep{Molloy:1995}. In this case, models
that incorporate heterogeneity like HetPW and PGF (which are numerically
indistinguishable in this case and several others) are in very close agreement
with simulation, while just taking the average degree as in HomPW is a poor
choice. In B, assortativity is added to the two group model following the
approach of~\citet{Newman02}, and HetPW outperforms PGF. Plots C and D show
regular graphs with four links per node, but while C is static in D the rate of
making and breaking links is much faster than the epidemic process. Models like
HomPW and PGF are therefore better for the former and Degree-based models are
better for the latter---in reality the ratio of the rate of network change to
the rate of transmision may not be either large or small and so a more
sophisticated method may be best~\citep{Eames:2004,Volz:2007}. E shows a graph
with four links per node where clustering has been introduced by the rewiring
method of~\citet{Bansal:2009} sometimes called the `big V'~\citep{House:2010a}.
In this case ClustPW performs better than HomPW and PGF, but clearly there is
significant inaccuracy around the region of peak prevalence and so this model
captures qualitatively the effects of clustering without appearing to be exact
for this precise network. Finally, Plot F considers the case of a
one-dimensional next-nearest-neighbour lattice (so there are four links per
node). This introduces long path lengths between nodes in addition to
clustering, meaning that the system does not converge onto a period of
asymptotic early growth and so realisations are shown as a density plot rather
than a confidence interval.  ClustPW accounts for clustering, but not long path
lengths and so is in poor agreement with simulation while the Pair-based curve
captures the qualitative behaviour of an epidemic on this lattice whilst being
quantitatively a reasonable approximation.

\section{Inference on Networks}

In order to be predictive, epidemic models rely on valid values for parameters
governing outbreak dynamics, conditional on the population structure.  However,
obtaining these parameters is complicated by the fact that, even when knowing
the underlying contact network structure, infection events are censored --- it
is only when disease is detected either from symptoms or laboratory tests that
a case becomes apparent.  In attempting to surmount this difficulty, parameter
estimates are often obtained by making strong assumptions as to the infectious
period, or through ad-hoc methods with unknown certainty.   Measuring the
uncertainty in such estimates is as important as obtaining the estimates
themselves in providing an honest risk prediction.  Given these difficulties,
inference for epidemic processes has perhaps received little attention in
comparison to its simulation counterpart. \\

The presence of contact network data for populations provides a unique
opportunity to estimate the importance of various modes of disease transmission
from disease incidence or contact tracing data.  For example, given knowledge
of the rate of contact between two individuals, it is possible to infer the
probability that a contact results in an infection.  If data on mere
connectivity (i.e.\ a 1 if the individuals are connected, and 0 otherwise) is
available, then it is still possible to infer a rate of infection between
connected individuals.  Thus the detail of the inference is determined to a
large extent by the available detail in the network data \citep{JewEtAl09c}.

\subsection{Availability of Data}

Epidemic models are defined in terms of times of transitions between infection
states, for example a progression from susceptible, to infected, to removed
(i.e. recovered with lifelong immunity or dead) in the so-called `SIR' model.
Statistical inference requires firstly that observations of the disease process
are made: at the very least this comprises the times of case detections,
remembering that infection times are always censored (you only ever know you
have a cold a few days after you caught it).  In addition, covariate data on
the individuals provides structure to the population, and begins to enable the
statistician to make statements about the importance of individuals'
relationships to one another in terms of disease transmission.  Therefore, any
covariate data, however slight, effectively implies a network structure upon
which disease transmission can be superimposed. \\

As long as populations are relatively small (e.g. populations of farms in
livestock disease analysis), it is common for models to operate at the
individual level, providing detailed information on case detection times and
perhaps even information on epidemiologically significant historical contact
events \citep{JewEtAl2009a, JewEtAl2009b, KeelEtAl01}.  In other populations,
however, such detailed data may not be available due to practical and ethical
reasons.  Instead, data is supplied on an aggregated spatial and/or temporal
basis.  For the purposes of inference, therefore, this can be regarded as a
household model, with areas constituting households. \\

In a heterogeneous population, the behaviour of an epidemic within any
particular locality is governed by the relationship between infected and
susceptible individuals.  For inference in the early stages of an epidemic, it
is important to quantify the amount of uncertainty in the underlying contact
networks as the early growth of the epidemic is known to be sub-exponential due
to the depletion of the local susceptible population. This contrasts markedly
to the exponential growth observed in a large homogeneously mixing population
\citep{AndersonMay}. When the network is known and details of individual
infections are available, contact tracing data may be used to infer the
network; this data could also be used for inference on the epidemic parameters
\citep{WallingaJ2006, Wallinga:2007}.  Conversely, if the network is completely
unknown, it would be useful if estimation of both the epidemic parameters and
parameters specifying the structure of the network was possible.  This is a
difficult problem because the observed epidemic contains very limited
information about the underlying network, as demonstrated by
\citet{britton2002bayesian}. However, with appropriate assumptions some results
can be obtained; the limited amount of existing work in this area is described
in section \ref{InfDisNet} below, although clearly the problem is worthy of
further study.

\subsubsection{Inference on Homogeneous Models}

For homogeneous models the basic reproduction number, or $R_{0}$, has several
equivalent definitions and can be defined in terms of the transmission rate
$\beta$ and removal rate $\gamma$.  For non-homogeneous models the definitions
are not equivalent, see for example \cite{L.Pellis11062009}. \\

Although inference for $\beta$ and $\gamma$ is difficult for real applications
(see below), it turns out that making inference on $R_{0}$ (as a function of
$\beta$ and $\gamma$) is rather more straight-forward. \citet{HefSmWah05}
summarise various methods for estimating $R_{0}$ from epidemiological data
based on endemic equilibrium, average age at infection, epidemic final size,
and intrinsic growth rate \citep{Mol95,DiekHees00,Het00}.  However, these
methods all rely on observing a complete epidemic, and hence for real-time
analysis during an epidemic we must make strong assumptions concerning the
number of currently undetected infections. An example of inference for $R_0$
based upon complete epidemic data is provided by \cite{stegeman}, where data
from the 2003 outbreak of High Pathogenicity Avian Influenza H7N7 is fitted to
a chain-binomial model using a generalised linear model. \\

Obviously, complete or near-complete epidemic data is rare and hence it is
desirable to perform inference based upon partial observation. This is
particularly relevant for real time estimation of $R_0$. For example,
\citet{CaucEtAl06} attempt to estimate $R_{0}$ in real-time by constructing a
discrete-time statistical model that imputes the number of secondary cases
generated by each primary case.  This is based on the method of
\citet{WalTeun04} who formulate a likelihood function for inferring who
infected whom from dates of symptom onset\[ L(i\mbox{ infected
}j)=\frac{w(t_{j}-t_{i})}{\sum_{j\ne k}w(t_{j}-t_{k})}\] where $w(\cdot)$ is
the probability density function for the \emph{generation interval}
$t_{j}-t_{i}$, i.e. the time between infector $i$'s infection time and infectee
$j$'s infection time. Of course, infection times are never observed in practice
so symptom onset times are used as a proxy, with the assumption that the
distribution of infection time to symptom onset time is the same for every
individual. Bayesian methods are used to infer ``late-onset'' cases from known
``early-onset'' cases, but large uncertainty of course remains when inferring
the reproductive ratio close to the current time as there exists large
uncertainty about the number of cases detected in the near future.
Additionally, a model for $w(\cdot)$ must be chosen
\citep[see][]{LipCohCoopRob03}. \\

The trade-off in the simplicity of estimating $R_{0}$ in these ways, however,
is that although a population-wide $R_{0}$ gives a measure of whether an
epidemic is under control on a wide-scale, it give no indication as to
regional-level, or even individual-level, risk. Moreover, the two examples
quoted above do not even attempt to include population heterogeneity into their
models, though the requirement for its inclusion is difficult to ascertain in
the absence of model diagnostics results.  It is postulated, therefore, that a
simple measure of $R_{0}$, although simple to obtain, is not sufficient in
order to make tactical control-policy decisions. In these situations, knowledge
of both the transmission rate {\em and} removal rate are required.

\subsection{Inference on Household Models}

Inference for households models is well developed in comparison to inference
for other `network' models. In essence this is for three main reasons: firstly,
it is a reasonable initial approximation to assume that infection either occurs
within the home or from a random source in the population. Secondly, entire
households can be serologically sampled following an epidemic, such that the
distribution of cases in households of given sizes can be ascertained. Finally,
it is often a reasonable approximation that following introduction of infection
into the household, the within-household epidemic will go extinct before any
further introductions --- which dramatically simplifies the mathematics.\\

The first methods proposed for such inference are maximum likelihood procedures
based upon chain-binomial models, such as the Reed-Frost model, or the
stochastic formulation of the Kermack-McKendrick model considered by
\citet{Bartlett:1949}.  These early methods are summarised by
\citet{Bailey:1975}. They, and the significant majority of methods proposed for
household inference to date, use final-size data which can be readily obtained
from household serology results. A simplifying assumption to facilitate
inference in most methods, is that the epidemics within the various households
evolve independently (e.g.~see the martingale method of \citet{Becker:1979},
which requires the duration of a latent period to be substantial for practical
implementation).

Additionally, fixed probabilities $p_C$ and $p_H$, corresponding to a
susceptible individual escaping community-acquired infection during the
epidemic and escaping infection when exposed to a single infected household
member, respectively, were initially assumed \citep{LonginiKoopman:1982}. Two
important, realistic extensions to this framework are to incorporate different
levels of risk factors for individuals \citep{LonginiKHC:1988} and to introduce
dependence of $p_H$ on an infectious period \citep{O'NeillBBEM:2000}. The
latter inclusion was enabled by appealing to results of \citet{BallMS:1997}.
These types of methods are largely based upon the ability to generate closed
form formulae for the final size distribution of the models. \\

The ability to relax assumptions further has been predominately due to use of
Markov chain Monte Carlo (MCMC) methods as first considered by
\citet{O'NeillBBEM:2000} for household models following earlier studies of
\citet{GibsonRenshaw:1998} and \citet{O'NeillRoberts:1999} who focused on
single, large outbreaks. This methodology has been used to in combination with
simulation and data augmentation approaches to tailor inference methods for
specific data sets of interest, e.g.~\citet{NealRob04} consider a model with a
spatial component of distance between households and data containing details of
dates of symptoms and appearance of rash, and has also resulted in a growing
number of novel methods for inference, e.g.~\citet{ClancyO'Neill:2007} consider
a rejection sampling procedure and \citet{CauchemezVBFF:2008} introduce a
constrained simulation approach. Even greater realism can be captured within
household models by considering the different compositions of households and
therefore the weighted nature of contacts within households. For example
\citet{Cauchemez:2004p6611} considered household data from the Epigrippe study
of influenza in France 1999--2000, and showed that children play a key role in
the transmission of influenza and the risk of bringing infection into the
household.\\

Whilst new developments are appearing at an increasing rate, the significant
majority of methods are based upon final size data and are developed for SIR
disease models, perhaps due in part to the simplification of arguments for
deriving final size distributions. One key, but still unanswered question from
these analyses of household epidemics is how the transmission rate between any
two individuals  in the household scales with the total number of individuals
in the household (compare \citet{LonginiKoopman:1982} and
\citet{Cauchemez:2004p6611}). Intuition would suggest that in larger households
the mixing between any two individuals is decreased, but the precise form of
this scaling is still unclear and much more data on large household sizes is
required to provide a definitive answer.\\

\subsubsection{Inference on Fully Heterogeneous populations}

Perhaps the holy grail of statistical inference on epidemics is to make use of
an individual-level model to describe heterogeneous populations at the limit of
granularity.  In this respect, Bayesian inference on stochastic mechanistic
models using MCMC have perhaps shown the most promise, allowing inference to be
made on both transmission parameters and using data augmentation to estimate
the infectious period. \\

An analysis of the 1861 outbreak of measles in Hagelloch by \citet{NealRob04}
demonstrates the use of a reversible jump MCMC algorithm to infer disease
transmission parameters and infectious period, whilst additionally allowing
formal comparisons to be made between several nested models. With the
uncertainty surrounding model choice, such methodology is vital to enable
accurate understanding and prediction. This approach has since been combined
with the algorithm of \citet{O'NeillRoberts:1999} and used to analyse disease
outbreaks such as avian influenza and foot and mouth disease in livestock
populations \citep{JewEtAl2009a,JewEtAl2009b,ChFer07}, and MRSA outbreaks in
hospital wards \citep{KypEtAl10}.\\

Whilst representing the cutting edge of inference on infectious disease
processes, these approaches are currently limited by computing power, with
their algorithms scaling by the number of infectives multiplied by the number
of susceptibles.  However, with advances in computer technology expected at an
increasing rate, and small approximations made in the calculation of the
statistical likelihoods needed in the MCMC algorithms, these techniques may
well form the mainstay of epidemic inference in the future.

\subsection{Inference From Contact Tracing}

In livestock diseases, part of the standard response to a case detection is to
gather contact tracing information from the farmer. The resulting data are a
list of contacts that have been made in and out of the infected farm during a
stipulated period prior to the notification of disease
\citep{defraExoticDiseaseFramework}. In terms of disease control on a local
level, this has the aim of identifying both the source of infection and any
presumed susceptibles that might have been infected as a result of the contact.
It has been shown that, providing the efficiency of following up any contacts
to look for signs of disease is high, this is a highly effective method of
slowing the spread of an epidemic, and finally containing it. \\

Much has been written on how contact tracing may be used to decrease the time
between infection and detection (notification) during epidemics.  However, this
focuses on the theoretical aspects of how contact tracing efficiency is related
to both epidemic dynamics and population structure (see for example
\citet{EamKeel03,KisGrKao05,KlFrHees06}). In contrast, the use of contact
tracing data in inferring epidemic dynamics does not appear to have been well
exploited, although it was used by the Ministry of Agriculture, Fisheries, and
Food (now Defra) to directly infer a spatial risk kernel for foot and mouth
disease in 2001. This assumed that the source of infection was correctly
identified by the field investigators, thereby giving an empirical estimate of
the probability of infection as a function of distance
\citep{FerDonAn01a,KeelEtAl01,Sav_et_al06}.  Strikingly, this shows a high
degree of similarity to spatial kernel estimates based on the statistical
techniques of \citet{Dig06} and \citet{Kyp07} without using contact tracing
information. However, \citet{CaucEtAl06} make the point that the analysis of
imperfect contact tracing data requires more complex statistical approaches,
although they abandoned contact tracing information altogether in their
analysis of the 2003 SARS epidemic in China.  Nevertheless, recent unpublished
work has shown promise in assimilating imperfect contact tracing data and case
detection times to greatly improve inference, and hence the predictive
capability of simulation techniques.

\subsubsection{Inference From Distributions Over Families of Networks}
\label{InfDisNet}

Qualitative results from simulations indicate that epidemics on networks, for
some parameter values, show features that distinguish them from homogeneous
models.  The principal features are a very variable length slow-growth phase,
followed by a rapid increase in the infection rate and a slower decline after
the peak \citep{Keeling:2005}.  However in quantitative terms there is usually
very limited information about the underlying network and parameters are often
not identifiable.  When the details of the network are unknown, but something
is known or assumed about its formation, estimation of both the epidemic
parameters and parameters for the network itself are in principle possible
using MCMC techniques.  All the stochastic models for generating networks
described in section \ref{ThNet} above realise a distribution over all or some
of the $2^{N(N-1)/2}$ possible networks.  In most cases this distribution is
not tractable; MCMC techniques are in principle still possible but in practice
would be too slow without careful design of algorithms.\\

However with appropriate assumptions some results can be obtained, which
provide some insight into what more could be achieved.  When the network is
taken to be an Erd\H{o}s-R\'{e}nyi graph with unknown parameter $p$ and the
epidemic is a Markovian SIR, \citet{britton2002bayesian} showed that it is
possible to estimate the parameters, although they highlight the ever-present
challenge of disentangling epidemiological from network parameters.  The MCMC
algorithm was improved by \citet{neal2005case} and the extension from SIR to
SEIR has been developed by \citet{Groendyke2010}.  However, the extension to
more realistic families of networks remains a challenging problem, and will
undoubtedly be the subject of exciting future research.

\section{Discussion}
The use of networks is clearly a rapidly growing field in epidemiology. By assessing (and quantifying) the potential transmission routes between individuals in a population, researchers are able to both better understand the observed distribution of infection as well as create better predictive models of future prevalence. We have shown how many of the structural features in commonly-used contact networks can be quantified and how there is an increasing understanding of how such features influence the propagation of infection. However, a variety of challenges remain.

\subsection{Open Questions}
Several open problems remain if networks are to continue to influence predictive epidemiology. The majority of these stem from the difficulty in obtaining realistic transmission networks for a range of pathogens. Although some work has been done to elucidate the interconnected structure of sexual encounters (and hence the sexual transmission network), these are still relatively small-scale compared to the population size and suffer from a range of potential biases. Determining comparable networks for airborne infections is a far greater challenge, due to the less precise definition of a potential contact.\\ 

One practical issue is therefore whether new techniques can be
developed that allow contact networks to be assessed
remotely. Proximity loggers, such as those used by Hamede and
colleagues \citep{Hamede:2009p7045}, provide one potential avenue
although it would require the technology to become sufficiently
robust, portable and cheap that a very large proportion of a
population could be convinced to carry one at all times. For many
human populations, where the use of mobile phones (which can detect
each other via Bluetooth) is sufficiently widespread, there is the
potential to use them to gather network information --- although the
challenges of developing sufficiently generic software should not be
underestimated. While these remotely sensed networks would provide
unparalleled information that could be obtained with the minimum of
effort, there would still be some uncertainty surrounding the nature
of each contact.\\

There is now a growing set of diary-based studies that have attempted to record the personal contacts of a large number of individuals; of these POLYMOD is currently the most comprehensive \citep{Mossong:2008p5450}. While such egocentric data obviously provides extensive information on individual behaviour, due to the anonymity of such surveys it is not clear how the alters should be connected together. The configuration method of randomly connecting half-links provides one potential solution, but what is ideally required is a more comprehensive method that would allow clustering, spatially localised connections and assortativity between degree distributions to be included and specified.\\

Associated with the desire to have realistic contact networks for entire populations, comes the need to characterise such networks in a relatively parsimonious manner that provides important insights into the types of epidemiological dynamics that could be realised. Such a characterisation would allow for different networks (from different times or different locations) to be compared in a manner that is epidemiologically significant; and would allow artificial networks to be created that matched particular known network features. This clearly relies on both existing measures of network structure (as outlined in section 3) together with a robust understand of how such features influence the transient epidemic dynamics (as outlined in section 4.2). However, such a generic understanding of all network features is unlikely to arise for many years. A more immediate challenge is to understand ways in which local network structure (clustering, cliques and spatially-localised connections) influence the epidemiological dynamics.\\

To date the vast majority of the work into disease transmission on networks has focused on static networks where all links are of equal strength and therefore associated with the same basic rate of transmission. However, it is clear that contact networks change over time (both on the short-time scale of who we meet each day, and on the longer time-scale of who our main work and social contacts are), and that links have different weights (such that some contacts are much more likely to lead to the transmission of infection than others). While the simulation of infection on such weighted time-varying networks is feasible, it is unclear how the existing sets of network properties or the existing literature of analytical approaches can be extended to such higher-dimensional networks.\\

For any methodology to have any substantive use in the field, it is important both to have effective data gathering protocols in place, and to have the statistical techniques in place to analyse it.  Here, three issues are perhaps most critical.  Firstly, data gathering resources are almost always limited.  Therefore, carefully designed randomised sampling schemata should be employed to maximise the power of the statistical techniques used to analyse data, rather than having to reply on data augmentation techniques to work around the problems present in ad-hoc datasets.   This aspect is particularly important when working on network data derived from population samples.  Secondly, any inference on both network and infectious disease models should be backed up by a careful analysis of model fit.  Although recent advances in statistical epidemiology have given us an unprecedented ability to measure population/disease dynamics based on readily available field data, epidemic model diagnostics are currently in their infancy in comparison to techniques in other areas of statistics.  Therefore it is expected that, with the growth in popularity of network models for analysing disease spread, much research effort will be required in designing such methodology.\\

\subsection{Conclusions}

We have highlighted that the study of contact networks is
fundamentally important to epidemiology and provides a wealth of tools
for understanding and predicting the spread of a range of
pathogens. As we have outlined above many challenges still exist, but
with growing interest in this highly interdisciplinary field and ever
increasing sophistication in the mathematical, statistical and
remote-sensing tools being used, these problems may soon be overcome.
We conclude therefore that now is an exciting time for research into
network epidemiology as many of the practical difficulties are
surmounted and theoretical concepts are translated into results of
applied importance in infection control and public health.\\

\section*{Acknowledgements}

This work is funded by the Medical Research Council (LD, MJK, MCV), the
Biotechnology and Biological Sciences Research Council (CPJ, MJK, GOR), the
Engineering and Physical Sciences Research Council (TH, AF, MJK), the Centre
for Research in Statistical Methodology funded by the Engineering and Physical
Sciences Research Council (GOR), the Research and Policy for Infectious Disease
Dynamics (RAPIDD) program of the Science and Technology Directorate (MJK) and
the Australian Research Council's Discovery Projects funding scheme, project
number DP110102893 (JVR). We would like to thank Kieran Sharkey for use of
pre-publication MATLAB code, and two anonymous reviewers for their comments.\\

\newpage
\section{Notation}

\begin{tabular}{|c|c|c|c|}
\hline
Concept/Measure & Other common names & Our notation & Other common notation \\
\hline
Network & Graph & $G$ & \\
Node & Vertex, point, site, actor &$n$ & $v$\\
Edge & Link, tie, bond & $l$ & $e$\\
Adjacency matrix & Connectivity matrix & $G_{ij}$ & $a_{ij},A_{ij}$\\
Number of nodes & Size of network & $N$ & $n, S$\\
Number of edges & Graph size & $L$ & $e$, $l$\\
Centrality & & $C$ & \\
Degree & Connectivity & $k$ & $d$, $C_d$\\
Betweenness & & $B_i$ & $\mbox{bet}_i$, $C_b$\\
Degree distribution & Connectivity distribution & $P(k)$ &$P_k$,$p_k$\\
Shortest path distance &Geodesic distance & $D_{i,j}$ & $d_{i,j}$\\
Clustering & transitivity & $\phi$ & $c$, $\Phi$ \\
Number of nodes of type A & & $[A]$ & $n_A$, $N_A$\\
Number of $A-B$ pairs & & $[AB]$ & $n_{AB}$, $N_{AB}$\\
Diameter & Maximal shortest path&$\mbox{Diam}(G)$ &$\mbox{max}(D_{i,j})$\\
\hline
\end{tabular}

\end{document}